\documentclass[aps,reprint,noshowkeys,superscriptaddress]{revtex4-1}
\usepackage{graphicx,dcolumn,bm,xcolor,microtype,multirow,amscd,amsmath,amssymb,amsfonts,physics,longtable,wrapfig,bbold,siunitx,xspace}
\usepackage[version=4]{mhchem}

\usepackage[utf8]{inputenc}
\usepackage[T1]{fontenc}
\usepackage{comment}

\usepackage{hyperref}
\hypersetup{
    colorlinks,
    linkcolor={red!50!black},
    citecolor={red!70!black},
    urlcolor={red!80!black}
}

\usepackage{balance}

\usepackage{listings}
\definecolor{codegreen}{rgb}{0.58,0.4,0.2}
\definecolor{codegray}{rgb}{0.5,0.5,0.5}
\definecolor{codepurple}{rgb}{0.25,0.35,0.55}
\definecolor{codeblue}{rgb}{0.30,0.60,0.8}
\definecolor{backcolour}{rgb}{0.98,0.98,0.98}
\definecolor{mygray}{rgb}{0.5,0.5,0.5}

\definecolor{sqred}{rgb}{0.85,0.1,0.1}
\definecolor{sqgreen}{rgb}{0.25,0.65,0.15}
\definecolor{sqorange}{rgb}{0.90,0.50,0.15}
\definecolor{sqblue}{rgb}{0.10,0.3,0.60}

\lstdefinestyle{mystyle}{
    backgroundcolor=\color{backcolour},
    commentstyle=\color{codegreen},
    keywordstyle=\color{codeblue},
    numberstyle=\tiny\color{codegray},
    stringstyle=\color{codepurple},
    basicstyle=\ttfamily\footnotesize,
    breakatwhitespace=false,
    breaklines=true,
    captionpos=b,
    keepspaces=true,
    numbers=left,
    numbersep=5pt,
    numberstyle=\ttfamily\tiny\color{mygray},
    showspaces=false,
    showstringspaces=false,
    showtabs=false,
    tabsize=2
  }

  \usepackage[]{notes2bib}
  \usepackage{siunitx}

  \newcolumntype{d}{D{.}{.}{-1}}

\usepackage{ulem}
\usepackage{booktabs} 
\usepackage{braket}
\usepackage{mathrsfs}
\usepackage{hyperref}
\usepackage{bm}
\usepackage{bbding}
\usepackage{pifont}
\newcommand{\LCT}{Sorbonne Universit\'e, CNRS, LCT UMR 7616, Paris 75005, France}
\newcommand{\MPI}{Max-Planck-Institut f\"ur Physik komplexer Systeme, N\"othnitzer Str. 38, 01187 Dresden, Germany}
\newcommand{\ICP}{Universit\'e Paris-Saclay, CNRS, Institut de Chimie Physique UMR8000, 91405, Orsay, France}

\begin{document}	
\title{Different perspectives on the exact factorization for photon-electron-nuclear systems}

\author{Claudia \surname{Magi}}
	\affiliation{\ICP}

\author{Peter \surname{Sch\"urger}}
        \email{pschuerger@pks.mpg.de}
	\affiliation{\MPI}

\author{David \surname{Lauvergnat}}
	\affiliation{\ICP}
	
\author{Federica \surname{Agostini}}
	\email{federica.agostini@sorbonne-universite.fr}
	\affiliation{\LCT}

\begin{abstract}
We employ the exact factorization of a multi-component wavefunction to analyze the dynamics of interacting photons, electrons and nuclei. We consider physical situations emerging in the regime of strong coupling between light excitations and molecular -- electronic -- excitations, giving rise to the so-called \textsl{molecular polaritons}. Nonadiabatic molecular dynamics techniques, routinely used in the field of chemical physics, have been often employed to simulate photophysical and photochemical phenomena in the presence of molecular polaritons. In this work, we analyze the foundations of these techniques in the eye of the exact factorization and we assess their performance on illustrative model studies.
\end{abstract}

\maketitle

\section{Introduction}\label{sec: intro}
Experimental realization of strong light-matter coupling and its consequences on physical and (photo)chemical properties of matter~\cite{Hutchison2012, Ebbesen_ACR2016, Rashidi2025, Garg2025} represented a stimulating motivation to introduce concepts of quantum optics in the fields of chemical physics and physical chemistry for novel theoretical developments.~\cite{Fregoni2018, Groenhof2024, Groenhof2022, Feist2018, Feist2022, Subotnik2022-2} In the regime where the rate of energy exchange between matter and light is faster than any dissipation effect, which can be achieved in confined spaces like optical or plasmonic cavities, matter excitations can hybridize with light excitations to yield new states of matter, so-called, polaritons.~\cite{Ebbesen_ACR2016} The theoretical formulation of this phenomenon requires to treat matter and light on equal, quantum-mechanical footing, giving rise to the field of cavity quantum electrodynamics (cQED).~\cite{Mandal2023,Flick2017_1,Ribeiro2018,Yuen-Zhou2026-tt, Ruggenthaler2023,Taylor2025, Rubio2018, Mukamel2023, Kowalewski2016}

A plethora of approaches have been reformulated or created for cQED, ranging from electronic structure theories,~\cite{Angelico2023, Riso2022, Haugland2020, Schfer2022, Wang2019, Ruggenthaler2014, Flick2017_2, Wickramasinghe2025, Bonini2022, Bonini2024} molecular dynamics and nonadiabatic molecular dynamics methods,~\cite{De2024,Krupp2025, Hoffmann2019, Hoffmann2019_jun, Maitra_JCP2020, Hu2023, Sokolovskii2024, Hu2025, Tichauer2021, Vendrell2018, Li2021, Li2021_2, Rana2023, Yu2022, Li2022} and phenomenological models.~\cite{Kowalewski2016, Galego2015, Galego2016, Galego2017} The challenges for these approaches are multiple. Firstly, the light degrees of freedom, i.e., the photons, become active players in polaritonic phenomena, rather than ``spectators'', as it is usually the case when treating light as an external classical field. Furthermore, accounting for many-body effects, collective effects, cavity losses, and long-range interactions is sometimes necessary but clearly stretches the capabilities of any theoretical formulation.~\cite{Li2021_2, Ghosh2025, Sokolovskii2024, Fregoni2018, Feist2022}.

This paper is devoted to address the first issue above, i.e., the description of the photons, by presenting an analysis of different perspectives in the context of nonadiabatic molecular dynamics (NAMD),~\cite{Curchod_WIRES2019} aiming to investigate with an \textsl{ab initio} viewpoint processes involving electronic excitations typically encountered in photophysics and photochemistry. Specifically, we employ the exact factorization of a multi-component wavefunction~\cite{Gross_PRL2010, Agostini_EPJB2021, Agostini_PCCP2024, Agostini_CPR2026} to formulate the quantum-mechanical photon-electron-nuclear (PEN) problem and to propose suitable approximations applicable in NAMD simulations. The exact factorization has been employed previously by various authors to explore its potential for applications in the context of polaritonic phenomena,~\cite{Maitra_EPJB2018, Tokatly_EPJB2018, Maitra_JCP2020, Maitra_JCP2021, Maitra_JCP2022, Maitra_PRL2019} and our group proposed recently a formulation that naturally lends itself to NAMD.~\cite{Agostini_JCP2024_2} In this work, we further develop this formulation of the exact PEN factorization to assess the performance of the approximations needed for the numerical treatment of the problem.

The exact factorization expresses a multi-component wavefunction as the product of a marginal amplitude and of a conditional amplitude. Such an expression allows one to reformulate the full time-dependent Schr\"odinger equation (TDSE) in terms of an effective TDSE for the marginal component, where the effect of the conditional enters as time-dependent scalar and vector potentials, and a non-Hermitian evolution equation for the conditional amplitude. The exact factorization has been employed extensively to treat electron-nuclear systems,~\cite{Gross_PRL2010, Agostini_EPJB2021, Agostini_PCCP2024} even though extensions to a variety of physical problems have been proposed.~\cite{GonZhoRei-EPJB-18, Maitra_PRL2020, Gross_PRL2021, Agostini_CPC2024, Agostini_JCP2025_DFT, CohSteGro-PRB-25, Hunter_IJQC1986, Schild_JPCL2021, Burghardt_PRL2024_GP, VM23, Franco_JCP2017, Suzuki_PRA2016, Scherrer_JCP2015, Scherrer_PRX2017, Blumberger_JCP2025}

The framework offered by the exact factorization is particularly well suited to analyze different aspects of PEN dynamics in the context of molecular polaritons, especially aiming to understand how the photonic degrees of freedom can be incorporated accurately and efficiently in the numerical solution of the TDSE. Previous studies have explored different, complementary formulations of the exact factorization applied in the field of molecular polaritons, and we focus in this work on two ``flavors'', with the aim to assess their suitability for the development of computational methods to solve the TDSE with \textsl{ab initio} NAMD techniques. These two flavors naturally arise from the different possibilities one faces when decomposing into marginal and conditional the PEN wavefunction. Nonetheless, the remarkable property of the exact factorization when employed in the context of PEN dynamics is that these two equivalent flavors can be introduced within the same formalism, even though they provide qualitatively different pictures of the same physical problem. Furthermore, we will investigate how differences arise in the approximate simulated dynamics, and we will draw direct connections to the approaches that have been extensively used in the literature.~\cite{Feist2018, Nitzan_PNAS2020}

The remainder of the paper is organized as follows. In Section~\ref{sec: theory}, we construct the exact-factorization framework by introducing these two flavors, which we refer to as \textsl{electronic perspective} and \textsl{polaritonic perspective}, and which will be discussed in detail all along the paper. In Section~\ref{sec: qc limit}, we discuss the implications of the two perspectives in the numerical treatment of PEN dynamics to describe the emergence of molecular polaritons. In Section~\ref{sec: results}, we report some numerical studies on illustrative models exemplifying (photochemical) nonadiabatic processes and (photophysical) Rabi oscillations. In Section~\ref{sec: conclusions}, we present our conclusions.

\section{Exact photon-electron-nuclear factorization}\label{sec: theory}
We consider a system of interacting photons, electrons and nuclei (PEN), which is described by the Hamiltonian
\begin{align}\label{eqn: PEN H}
    \hat{H}(\mathbf{r},\mathbf{q},\mathbf{R}) 
    = \hat{H}_M(\mathbf{r},\mathbf{R}) 
      +  \hat{H}_P(\mathbf{q})  + \hat{H}_{PM}(\mathbf{r},\mathbf{q},\mathbf{R}) 
\end{align}
This Hamiltonian is the Pauli-Fierz non-relativistic QED Hamiltonian in the dipole gauge~\cite{Mandal2023,Woolley2020,Taylor2025} and can be derived from the minimal coupling Hamiltonian in the Coulomb gauge by applying the Power-Zienau-Woolley Gauge transformation~\cite{Power1959} and a unitary phase transformation assuming that the dimension of the system is much smaller than the length of the cavity~[\citenum{Mandal2020}].

In Eq.~(\ref{eqn: PEN H}), we indicate electronic coordinates with the symbol $\mathbf{r}$, nuclear coordinates with $\mathbf{R}$ and the photon displacement with $\mathbf{q}$. $\hat{H}_M$ is the matter Hamiltonian, thus it only depends on the matter degrees of freedom, i.e., $\mathbf{r}$ and $\mathbf{R}$. It contains the nuclear kinetic energy and the electronic Hamiltonian, often referred to as ``Born-Oppenheimer'' ($BO$) Hamiltonian,
\begin{align}\label{eqn: matter H}
    \hat{H}_M(\mathbf{r},\mathbf{R}) &= \hat T_n(\mathbf R)+ \hat H_{BO}(\mathbf r,\mathbf R)\\
    &=\sum_{\alpha=1}^{N_n} \frac{-\hbar^2\boldsymbol\nabla_\alpha^2}{2M_\alpha} + \hat H_{BO}(\mathbf r,\mathbf R)
\end{align}
Here, the index $\alpha$ runs over $N_n$ nuclei, $\boldsymbol\nabla_\alpha$ is the gradient with respect to the position of the nucleus $\alpha$ and $M_\alpha$ is the corresponding mass. Quantization of the electromagnetic field yields the photon Hamiltonian $\hat{H}_P$, which is the sum of harmonic oscillators
\begin{align}\label{eqn: photon H}
        \hat{H}_P(\mathbf{q})
    &= \hat{T}_P(\mathbf{q}) + \hat{V}_P(\mathbf{q}) \\
    &= \sum_{\nu=1}^{2N_P} \frac{1}{2} 
       \left( \hat{p}_\nu^2 + \omega_\nu^2 \hat{q}_\nu^2 \right)=\sum_{\nu=1}^{2N_P} \frac{1}{2} 
       \left(-\frac{\partial^2}{\partial q_\nu^2} + \omega_\nu^2 \hat{q}_\nu^2 \right)
\end{align}
with unitary masses and frequencies $\omega_\nu$. The sum over $\nu$ runs over both possible transverse polarizations of the electromagnetic field, i.e., $2N_P$. The photonic degrees of freedom are expressed in terms of the displacement operator $\hat q_\nu=\sqrt{\hbar/(2\omega_\nu)}(\hat a_\nu^\dagger+\hat a_\nu)$ and by its conjugated momentum operator $\hat p_\nu=i\sqrt{\hbar/(2\omega_\nu)}(\hat a^\dagger_\nu-\hat a_\nu)$, which are related to the electric field and to the magnetic field, respectively, using the annihilation and creation operators $\hat a_\nu,\hat a^\dagger_\nu$. Finally, the photon-matter coupling Hamiltonian $\hat H_{PM}$ in the dipole approximation is
\begin{align}\label{eqn: photon-matter H}
    \hat{H}_{PM}(\mathbf{r},\mathbf{q},\mathbf{R}) = \sum_{\nu=1}^{2N_P}\omega_\nu g_\nu\hat{q}_\nu \left(\sum_{\alpha=1}^{N_n}Z_\alpha e\hat{\mathbf R}_\alpha-\sum_{i=1}^{N_e}e\hat{\mathbf r}_i\right)
\end{align}
where $g_\nu$ is the coupling constant between the matter and the photons, $Z_\alpha e$ is the charge of the nucleus $\alpha$ (with $e$ the elementary charge) at position $\mathbf R_\alpha$, and $\hat{\mathbf r}_i$ is the electronic position operator ($N_e$ is the number of electrons). Note that in the above equation, we neglected the self-polarization term, which depends only on the matter degrees of freedom.~\cite{Maitra_JCP2020,Tichauer2021,Agostini_JCP2024_2}

The time evolution of the interacting PEN system is dictated by the TDSE
\begin{align}\label{eqn: PEN TDSE}
    i\hbar \frac{\partial}{\partial t}\Psi(\mathbf{r},\mathbf{q},\mathbf{R}, t) = \hat{H}(\mathbf{r},\mathbf{q},\mathbf{R}) \Psi(\mathbf{r},\mathbf{q},\mathbf{R}, t)
\end{align}
where $\Psi(\mathbf{r},\mathbf{q},\mathbf{R}, t)$ is the PEN time-dependent wavefunction. Employing the idea of the exact factorization,~\cite{Agostini_PCCP2024} such a multi-component wavefunction can be expressed as the product of a marginal and a conditional amplitude.
In this work, we propose two alternative flavors for such a decomposition
\begin{align}
    \Psi(\mathbf{r},\mathbf{q},\mathbf {R},t)&=\chi^{\mathrm{(EL)}}(\mathbf{q},\mathbf{R},t)\Phi^{\mathrm{(EL)}}(\mathbf{r},t;\mathbf{q},\mathbf{R})\label{el-eq}\\
        &=\chi^{\mathrm{(POL)}}(\mathbf{R},t)\Phi^{\mathrm{(POL)}}(\mathbf{r},\mathbf{q},t;\mathbf{R}) \label{pol-eq}
\end{align}
which will provide an \textsl{electronic perspective} and a \textsl{polaritonic perspective}, respectively, on the PEN dynamics. Note that we referred to this electronic perspective as ``cavity exact factorization'' in Ref.~[\citenum{Agostini_CPR2026}], thus using a definition related to the more general literature on cQED.~\cite{Fischer2023, Flick2017_1, Flick2017_2, Kowalewski2016, Wickramasinghe2025, Bonini2022, Bonini2024,Angelico2023} In Eq.~\eqref{el-eq}, the conditional component is an electronic conditional amplitude, that depends parametrically on the nuclear positions and on the photonic displacement, whereas in Eq.~\eqref{pol-eq}, the conditional term has a ``mixed'' electronic and photonic character, and depends parametrically on the nuclear positions.

In the sections that follow, we will discuss in detail these two perspectives, specifically analyzing the picture of the coupled PEN dynamics that they convey, also in relation to approximate NAMD schemes to simulate such dynamics. Also, we will draw connections to existing approaches that have been used in the context of molecular polaritons. Before moving to that, though, let us present the general form of the evolution equations driving the dynamics of the marginal and conditional amplitudes, highlighting the fact that they are formally identical using both perspectives.

Labeling with the symbol $\bm{X}$ the dependence of the marginal amplitude and with $\bm{x}$ the remaining degrees of freedom, the evolution equations for the marginal $\chi(\bm{X},t)$ and for the conditional $\Phi(\bm x,t;\bm X)$ amplitudes are
\begin{widetext}
\begin{align}
i\hbar \frac{\partial}{\partial t} \chi(\bm X,t) &= \left[\sum_{\Gamma} \frac{[-i\hbar \boldsymbol{\nabla}_\Gamma+\mathbf A_\Gamma(\bm X,t)]^2}{2M_\Gamma} + \epsilon(\bm X,t)\right]\chi(\bm X,t)\label{eqn: marginal eqn}\\
i\hbar \frac{\partial}{\partial t} \Phi(\bm x,t; \bm X) &= \left[\hat H_C (\bm x, \bm X)+\hat U_{coup}[\chi,\Phi]-\epsilon(\bm X,t)\right]\Phi(\bm x,t; \bm X)\label{eqn: conditional eqn}
\end{align}
\end{widetext}
For the electronic perspective, $\Gamma = \alpha,\nu$ and the masses are either the nuclear masses or unitary (for the photons' degrees of freedom); for the polaritonic perspective, $\Gamma = \alpha$ and the masses are only the nuclear ones. The ``conditional Hamiltonian'' $\hat H_C (\bm x, \bm X)$ is either $\hat H-(\hat T_n+\hat T_P)$ in the electronic perspective or $\hat H-\hat T_n$ in the polaritonic perspective. The time-dependent potential energy surface (TDPES)~\cite{Agostini_JPCA2022, Gross_JCP2015}
\begin{align}\label{eqn: tdpes}
    \epsilon(\bm X,t) = \left\langle \Phi(t;\bm X)\left|\hat H_C + \hat U_{coup} -i\hbar\frac{\partial}{\partial t}\right| \Phi(t;\bm X)\right\rangle_{\bm x}
\end{align}
and the time-dependent vector potential (TDVP)~\cite{Agostini_JPCL2017}
\begin{align}\label{eqn: tdvp}
    \mathbf A_\Gamma(\bm X,t)=\left\langle \Phi(t;\bm X)\big|-i\hbar \boldsymbol{\nabla}_\Gamma\Phi(t;\bm X)\right\rangle_{\bm x}
\end{align}
mediate the coupling between $\bm X$ and $\bm x$, along with the coupling operator
\begin{widetext}
\begin{align}\label{eqn: coup}
    \hat U_{coup}&[\chi,\Phi] =\sum_\Gamma \frac{[-i\hbar\boldsymbol{\nabla}_\Gamma - \mathbf A_\Gamma (\bm X,t)]^2}{2M_\Gamma}+\sum_\Gamma \frac{1}{M_\Gamma} \left(\frac{-i\hbar \boldsymbol{\nabla}\chi(\bm X,t)}{\chi(\bm X,t)}+\mathbf A_\Gamma (\bm X,t)\right)\cdot \left(-i\hbar\boldsymbol{\nabla}_\Gamma - \mathbf A_\Gamma (\bm X,t)\right)
\end{align}
\end{widetext}
The symbol $\langle\,\,\cdot\,\,\rangle_{\bm x}$ stands for an integration over $\bm x$, and we used the convention that the variable which is integrated out does not appear within the bra-ket.

Equations~\eqref{eqn: marginal eqn} to~\eqref{eqn: coup} are the usual equations of the exact factorization, which hold valid independently of the perspective employed to study the multi-component PEN dynamics. While adopting one or the other perspective is completely equivalent at the quantum-mechanical level, the performance of approximate schemes to simulate the coupled dynamics vary from one formulation to the other, as we will demonstrate below.

In Sections~\ref{sec: el p} and~\ref{sec: pol p}, we will introduce the electronic and the polaritonic perspectives, along with the idea of solving the PEN dynamics using NAMD techniques. To this end, we will refer to the \textsl{quantum-classical} treatment of the problem: classical-like trajectories are used to mimic the marginal degrees of freedom coupled to the quantum dynamics of the conditional degrees of freedom. Later on, in Section~\ref{sec: qc limit}, we will present the equations derived in the context of the exact factorization to simulate the NAMD of a PEN system using coupled trajectories, and we will draw connections to other NAMD schemes based on independent trajectories.

\subsection{Electronic perspective}\label{sec: el p}
The electronic perspective has been proposed previously in Ref.~[\citenum{Agostini_JCP2024_2}] in the context of PEN dynamics, but earlier work by Maitra and coworkers analyzed analogous problems by considering the photon displacement as the marginal variable.~\cite{Maitra_EPJB2018, Maitra_JCP2022}

Adopting the electronic perspective, the PEN dynamics is visualized in terms of a photon-nuclear wavefunction that evolves in time under the effect of the electrons, which appear in the problem via the electronic time-dependent potentials of Eqs.~\eqref{eqn: tdpes} and~\eqref{eqn: tdvp}. Aiming to simulate such dynamics using NAMD techniques, the ``usual'' strategy would be (1) to approximate the evolution of the marginal photon-nuclear wavefunction in terms of classical-like trajectories~\cite{Gross_JCTC2016, Ciccotti_EPJB2018, Ciccotti_JPCA2020} and (2) to express the electronic evolution in terms of properties such as energies, gradients and couplings, emanating from a Born-Huang-like representation. In this spirit, the electronic conditional amplitude is expanded as
\begin{align}\label{eqn: BH in electronic}
   \Phi^{\mathrm{(EL)}}(\mathbf{r},t;\mathbf{q},\mathbf{R})
    = \sum_{k=1}^{N_{st}} C_k^{\mathrm{(EL)}}(\mathbf{q},\mathbf{R},t)\,\phi_k(\mathbf{r};\mathbf{q},\mathbf{R})
\end{align}
with the stationary electronic Schr\"odinger equation
\begin{align}\label{eqn: def EEL}
    \left[\hat H-\hat T_n-\hat T_P\right]\,\phi_k(\mathbf{r};\mathbf{q},\mathbf{R}) =E_k^{\mathrm{(EL)}}(\mathbf{q},\mathbf{R})\,\phi_k(\mathbf{r};\mathbf{q},\mathbf{R})
\end{align}
defining the electronic eigenstates $\phi_k(\mathbf{r};\mathbf{q},\mathbf{R})$ and electronic potential energy surfaces $E_k^{\mathrm{(EL)}}(\mathbf{q},\mathbf{R})$, that depend on the nuclear positions and on the photon displacements. We use the index $k$ to label the $N_{st}$ electronic states considered in the expansion~\eqref{eqn: BH in electronic}.

In a NAMD procedure derived from the exact factorization, an ensemble of $N_{tr}$ trajectories representing the nuclei and the photonic displacements, $\bm X^{(I)}_\Gamma(t) = \mathbf R_\alpha^{(I)}(t),\mathbf q_\nu^{(I)}(t)$, moves according to a Newton-like equation with force $\mathbf F_\Gamma^{(I)}(t)$ encoding the effect of the TDPES and TDVP. In practice, the TDPES and TDVP are constructed by inserting Eq.~\eqref{eqn: BH in electronic} into Eqs.~\eqref{eqn: tdpes} and~\eqref{eqn: tdvp}, such that the necessary electronic-structure properties are readily available in any quantum chemistry package.~\cite{Gross_JPCL2017, Agostini_JCTC2024}

When compared to the more standard electron-nuclear problem usually treated with NAMD, the dimensionality of the nuclear problem is only increased by the inclusion of the photons' degrees of freedom. Since the photon-nuclear degrees of freedom are represented in terms of trajectories, the computational cost remains reasonable. Nonetheless, the dependence of the electronic eigenstates on the additional photonic coordinates needs to be explicitly taken into account.

In the electronic perspective, one can indeed question the appropriateness of treating the light photons' degrees of freedom using classical trajectories.~\cite{Scribano_JCTC2022} Many strategies have been devised in the literature aiming to account for quantum-mechanical effects in combination with a trajectory-based description of light nuclei, like protons, which may represent potential avenues for the developments of NAMD schemes adopting the electronic perspective in the exact factorization.~\cite{Rassolov_JCTC2023, Suzuki_PRA2016, Ananth_JCP2013, Huo_JCP2017} In addition, we will demonstrate in Section~\ref{sec: results}, that the performance of this perspective, even in comparison to the polaritonic perspective, depends on the studied physical situation. Finally, we note that this electronic perspective has been explored by other authors in the context of vibrational strong coupling.~\cite{Feist2018, Nitzan_PNAS2020}

\subsection{Polaritonic perspective}\label{sec: pol p}
The polaritonic perspective arising in the context of the exact factorization allows one to draw direct connections to the abundant literature that has been devoted to the theoretical treatment of molecular polaritons in the regime of electronic strong coupling.~\cite{Schfer2018, De2024, Fregoni2020, Hu2023, Sokolovskii2024, Hu2025} In this case, the PEN dynamics can be visualized in a way that is more familiar to NAMD, namely in terms of nuclei evolving in time guided by a ``modified'' electronic energy landscape, where the bare electronic effects are altered by the hybridization with the photons. Using the exact factorization, the marginal nuclear wavefunction thus evolves under the effect of the photon-electronic, i.e., polaritonic, time-dependent potentials.

When aiming at formulating the problem for NAMD, one needs (1) to approximate the evolution of the marginal nuclear wavefunction in terms of classical-like trajectories and (2) to express the polaritonic evolution in terms of properties such as energies, gradients and couplings, emanating from a Born-Huang-like representation. These properties are not \textsl{bare electronic structure properties} but have to be ``dressed'' by the light field. Therefore, the conditional amplitude is expanded as
\begin{align}\label{eqn: BH in polaritonic}
   \Phi^{\mathrm{(POL)}}(\mathbf{r},\mathbf{q},t;\mathbf{R})
    = \sum_{k=1}^{\tilde N_{st}} C_k^{\mathrm{(POL)}}(\mathbf{R},t)\,\varphi_k(\mathbf{r},\mathbf{q};\mathbf{R})
\end{align}
with the stationary polaritonic Schr\"odinger equation
\begin{align}\label{eqn: def EPOL}
    \left[\hat H-\hat T_n\right]\,\varphi_k(\mathbf{r},\mathbf{q};\mathbf{R}) =E_k^{\mathrm{(POL)}}(\mathbf{R})\,\varphi_k(\mathbf{r},\mathbf{q};\mathbf{R})
\end{align}
defining the polaritonic eigenstates $\varphi_k(\mathbf{r};\mathbf{q},\mathbf{R})$ and polaritonic potential energy surfaces $E_k^{\mathrm{(POL)}}(\mathbf{R})$, that depend only on the nuclear positions. In this case, the expansion in Eq.~\eqref{eqn: BH in polaritonic} includes $\tilde N_{st}\neq N_{st}$ terms.

Similarly to the discussion that we presented in Section~\ref{sec: el p}, a NAMD procedure derived from the exact factorization and employing the polaritonic perspective relies on the representation of the nuclear dynamics in terms of $N_{tr}$ trajectories, $\bm X_\Gamma^{(I)}(t)=\mathbf R_\alpha^{(I)}(t)$, which are generated from Newton's equation with a force $\mathbf F_\Gamma^{(I)}(t)$ that encodes the effect of the TDPES and TDVP. In turn, the TDPES and TDVP are constructed by inserting Eq.~\eqref{eqn: BH in polaritonic} into Eqs.~\eqref{eqn: tdpes} and~\eqref{eqn: tdvp}. In this case, however, the polaritonic Hilbert space is larger than the bare electronic Hilbert space, since it is the vector product of the electronic and photonic spaces. In the context of \textsl{ab initio} NAMD, usually the computational bottleneck is represented by the electronic structure problem, thus the growth of the number of states might become a severe limitation for the application of the polaritonic perspective, depending on the NAMD scheme employed.

\begin{figure}
    \centering
    \includegraphics[width=0.9\linewidth]{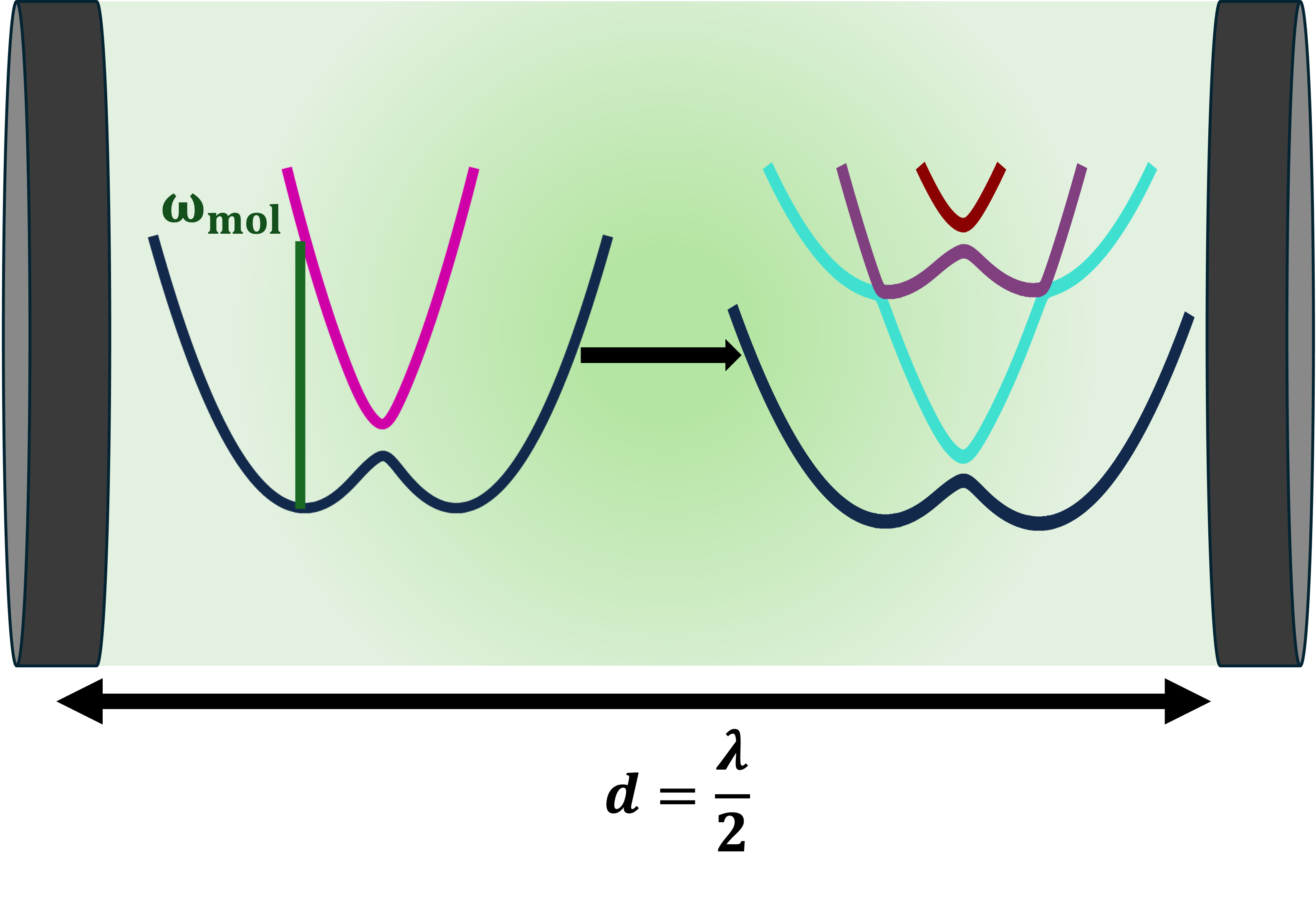}
    \caption{Schematic representation of a single mode cavity with a maximum of one photon, represented by the two mirrors at a distance $d = \lambda/2$, and its effect on a two-level system.}
    \label{fig: scheme}
\end{figure}

Figure~\ref{fig: scheme} schematically shows the effect of the strong coupling between a two-level system and a single mode cavity with a maximum of one photon, visualized in the polaritonic perspective. The cavity is represented by the two mirrors at a distance $d = \lambda/2$ with $\lambda = 2\pi/\omega_{mol}$. The two-level system is, instead, represented by the two electronic potential energy curves on the left, with the ground state having a double-well shape and forming an avoided crossing with the excited state, where the energies are close to each other. The potentials are plotted as functions of a one-dimensional nuclear coordinate, e.g., a reaction coordinate representing the progression of the reaction. When the system is strongly coupled with a single cavity mode in resonance with $\omega_{mol}$, the potential energy landscape is altered from the left scheme to the right one. Firstly, we observe that the number of states is doubled, since the light system can be in the zero-photon and in the one-photon state. Secondly, new avoided crossings are formed by the hybridization of the excitation of the system with the light excitation. The formation of new states and new avoided crossings affects the behavior of the nuclei, thus altering the dynamics from the cavity-free situation (left) to the strong coupling regime (right).

The polaritonic perspective has been extensively employed in the domain of molecular polaritons, even beyond the formulation of the problem based on the exact factorization. Therefore, we find particularly instructive in this work to compare this formulation in the context of NAMD with the electronic perspective, using the same formalism, the same approximations and the same software to simulate prototypical model situations in the regime of strong light-matter coupling.~\cite{Rana2023, Fregoni2020, Krupp2025, Tichauer2021, Groenhof2024}

\section{Nonadiabatic molecular dynamics with strong light-matter coupling}\label{sec: qc limit}
In this section, we introduce the quantum-classical treatment of PEN dynamics that we will use for \textsl{ab initio} NAMD simulations in Section~\ref{sec: results}. To this end, we will first describe how to perform the classical limit of the marginal evolution equation of the exact factorization, leading to the derivation of the coupled-trajectory mixed quantum-classical (CTMQC) algorithm.~\cite{Gross_PRL2015} Afterwards, we will extend this idea beyond the exact factorization, describing additional methodologies widely used in the field of \textsl{ab initio} NAMD, namely the independent-trajectory techniques known as multi-trajectory Ehrenfest (MTE)~\cite{TULLY1998} and Tully surface hopping (TSH).~\cite{Tully_JCP1990} For more details about the algorithms, we refer to Appendix~\ref{app: algorithms}. In addition, we will provide here only some general guidelines explaining how the classical treatment of the marginal wavefunction is introduced, without entering in the details of the approximations. For this, we refer the interested reader to Refs.~[\citenum{Agostini_PCCP2024, Agostini_JCTC2020_1, Gross_JCTC2016, Agostini_EPJB2021, IBELE2024188}]. 

Henceforth, we will use the general notation employed also in Eqs.~\eqref{eqn: marginal eqn} and~\eqref{eqn: conditional eqn}, namely $\bm X$ for the marginal variables and $\bm x$ for the remaining, conditional variables.

The quantum-classical reformulation of the exact factorization requires to introduce the concept of trajectories to mimic the evolution of the marginal wavefunction. This is formally achieved by replacing in the equations the variable $\bm X$ with the trajectories $\bm X^{(I)}(t)$, where the index $I=1,\dots, N_{tr}$ labels the trajectories of the ensemble, supposing that only an ensemble of trajectories is able to capture some aspects of the behavior of the marginal wavefunction, as for instance its delocalization in configuration space. For each $I$ and at time $t$, the $\bm X^{(I)}(t)$ has the same dimension as $\bm X$. Each trajectory is associated to a velocity $\dot{\bm X}^{(I)}(t)$ and is generated via Newton's equation, $M_\Gamma \ddot{\bm X}^{(I)}_\Gamma (t)= \mathbf F^{(I)}_\Gamma(t)$. Here, we recall that the index $\Gamma$ labels the particles described by the marginal wavefunction, such that both $\ddot{\bm X}^{(I)}_\Gamma (t)$ and $\mathbf F^{(I)}_\Gamma(t)$ are three-dimensional vectors in a general -- molecular -- application.

The effective Hamiltonian in Eq.~\eqref{eqn: marginal eqn} can be easily interpreted as a classical Hamiltonian function featuring time-dependent scalar and vector potentials, from which the classical force $\mathbf F^{(I)}_\Gamma(t)$ can be derived by simple application of the gradient $\boldsymbol{\nabla}_\Gamma$. Nonetheless, a more rigorous derivation has been proposed previously, employing the polar representation of the marginal wavefunction to obtain a Hamilton-Jacobi-like equation which is solved with the method of characteristics. Also in this case, details can be found in Refs.~[\citenum{Ciccotti_EPJB2018, Ciccotti_JPCA2020}].

The trajectories generated by the force $\mathbf F^{(I)}_\Gamma(t)$ can be assimilated to a moving grid, and the value of a generic function $f(\bm X,t)$ in the quantum-mechanical formulation can be recovered as $f(\bm X^{(I)}(t),t)$ in the quantum-classical formulation. For a very large number of trajectories $N_{tr}$, all configuration space is sampled at all times such that information is not lost when going from the quantum to the quantum-classical treatment, under the assumption that the trajectories explore sufficiently well the configuration space, similarly to the wavefunction. It is important to note that, in order to evaluate how functions of the type $f(\bm X^{(I)}(t),t)$ evolve, only total time-derivatives can be computed instead of partial time-derivatives.

This observation is crucial when accounting for the classical limit also in the conditional equation~\eqref{eqn: conditional eqn}, which has to be reformulated to accommodate for such a classical approximation. First, similarly to the discussions in Sections~\ref{sec: el p} and~\ref{sec: pol p}, we insert Eq.~\eqref{eqn: BH in electronic} or Eq.~\eqref{eqn: BH in polaritonic} in Eq.~\eqref{eqn: conditional eqn}. Then, we derive evolution equations for the expansion coefficients, thus either $\dot C_k^{\mathrm{(EL)}}(\bm X^{(I)}(t),t)$ or $\dot C_k^{\mathrm{(POL)}}(\bm X^{(I)}(t),t)$, depending on the perspective. Below, we will use the general symbol $\dot C_k^{(I)}(t)$, indicating the full time derivative of the coefficient $C_k^{(I)}(t)$ along the trajectory $I$ at time $t$.

Following the general guidelines described so far, the CTMQC algorithm can be summarized as the equations
\begin{align}
    \mathbf F^{(I)}_\Gamma(t) &= \mathbf F^{(I)}_{\Gamma,\mathrm{mf}}(t) + \mathbf F^{(I)}_{\Gamma,\mathrm{ct}}(t)\label{eqn: F in CTMQC}\\
    \dot C_k^{(I)}(t) &= \dot C_{k,\mathrm{mf}}^{(I)}(t)+\dot C_{k,\mathrm{ct}}^{(I)}(t)\label{eqn: Cdot in CTMQC}
\end{align}
dictating the evolution of the trajectories that mimic the time-dependence of the marginal amplitude (Eq.~\eqref{eqn: F in CTMQC}) and the evolution of the expansion coefficients to reconstruct the conditional amplitude in time (Eq.~\eqref{eqn: Cdot in CTMQC}). In each equation, the first term (mf) is a mean-field term and the second term (ct) is a coupled-trajectory term (basically arising from the dependence of the coupling operator~\eqref{eqn: coup} on the marginal wavefunction). Detailed derivations of these equations have been discussed previously, for instance in Refs.~[\citenum{Agostini_PCCP2024, Agostini_JCTC2020_1, Gross_JCTC2016, Agostini_EPJB2021, IBELE2024188}], and we report in Appendix~\ref{app: algorithms} the explicit expressions of the mean-field and coupled-trajectory terms.

By simply neglecting the coupled-trajectory terms in Eqs.~\eqref{eqn: F in CTMQC} and~\eqref{eqn: Cdot in CTMQC}, we recover the mean-field approximation, underlying the MTE scheme. In MTE, the coupling between the marginal and the conditional degrees of freedom is, thus, described in a mean-field way employing independent trajectories. Further simplifications of the classical force yield, instead, the TSH method. Specifically, in TSH the trajectories representing the evolution of the marginal amplitude are simply propagated -- also independently -- under the effect of forces obtained from the energy eigenvalues in Eq.~\eqref{eqn: def EEL} or in Eq.~\eqref{eqn: def EPOL} such that $\mathbf F^{(I)}_\Gamma(t)=-\boldsymbol{\nabla}_\Gamma E_{\mathrm a}(\bm X^{(I)}(t)) = \mathbf F^{(I)}_{\Gamma,\mathrm a}(t)$. Here, we removed the superscript ``(EL)'' or ``(POL)'' to highlight that this expression is valid when employing the electronic and the polaritonic perspective, and we used the subscript ``a'' to indicate that, at each time, the force is determined from the ``active'' (electronic or polaritonic) state. Such active state is selected stochastically at each time, allowing thus the trajectories to \textsl{hop} from one state to another in course of the dynamics.

Table~\ref{tab: summary methods} summarizes the NAMD methodologies introduced in this section, indicating the basic equations and their independent/coupled-trajectory character.
\begin{table}[h]
\small
  \caption{\ Summary table of the NAMD methodologies introduced in Section~\ref{sec: qc limit}.}
  \label{tab: summary methods}
  \begin{tabular*}{0.48\textwidth}{@{\extracolsep{\fill}}c|ccc}
    \hline
     & CTMQC & MTE & TSH \\
    \hline
    \hline
    Force & \phantom{$\bigg|$} $\mathbf F^{(I)}_{\Gamma,\mathrm{mf}}+\mathbf F^{(I)}_{\Gamma,\mathrm{ct}}$ & $\mathbf F^{(I)}_{\Gamma,\mathrm{mf}}$ & $\mathbf F^{(I)}_{\Gamma,\mathrm a}$ \\
    \hline
    Coefficients & \phantom{$\bigg|$} $\dot C_{k,\mathrm{mf}}^{(I)}+\dot C_{k,\mathrm{ct}}^{(I)}$ & $\dot C_{k,\mathrm{mf}}^{(I)}$ & $\dot C_{k,\mathrm{mf}}^{(I)}$ \\
    \hline
    Coupled trajectories & \phantom{$\bigg|$} \ding{51} & \ding{55} & \ding{55} \\
    \hline
  \end{tabular*}
\end{table}

We will employ these NAMD techniques to simulate the dynamics in a PEN system in the regime of strong light-matter coupling and we will analyze various physical conditions adopting the electronic perspective and the polaritonic perspective. The comparisons among the numerical results will allow us to assess the performance of each technique also in relation to the ``most suitable'' perspective.

\section{Illustrative model studies}\label{sec: results}
The electron-nuclear system that we will use as illustrative example features two electronic states, labeled S$_0$ and S$_1$, coupled to a one-dimensional nuclear mode identified by the variable $R$. In order to create two different physical situations, we will tune the parameters of the Hamiltonian (given below) such that, in one case, we model a photochemical reaction in the presence of nonadiabatic effects, showing how the strong coupling with a single mode cavity alters the progression of the reaction and, in the other case, we observe the emergence of Rabi oscillations attesting to the resonant energy exchange between the electronic excitation of the system and the light field.~\cite{Agostini_JCP2024_2}

The Hamiltonian is given in the diabatic electronic basis, such that $\hat H_M$ is the sum of the nuclear kinetic energy $\frac{-\hbar^2}{2M}\frac{\partial^2}{\partial R^2}$ and of
\begin{align}
    \hat H_{BO}(R) = \left(
    \begin{array}{cc}
        \frac{1}{2}k(R-R_1^2) & b\,e^{-a(R-R_3)^2} \\
        b\,e^{-a(R-R_3)^2} & \frac{1}{2}k(R-R_2^2)+\Delta 
    \end{array}
    \right)
\end{align}
the photon Hamiltonian is the sum of the kinetic energy term $-\frac{1}{2}\frac{\partial^2}{\partial q^2}$ and of 
\begin{align}
    \hat V_{P}(q) = \left(
    \begin{array}{cc}
        \frac{1}{2}\omega^2q^2 & 0 \\
        0 & \frac{1}{2}\omega^2q^2
    \end{array}
    \right)
\end{align}
and the photon-matter interaction Hamiltonian is modeled as
\begin{align}
    \hat H_{PM}(q,R) = \left(
    \begin{array}{cc}
        \omega \,g \,q\, ZR & \omega \,g \, q\, \mu \\
        \omega \,g \,q\,\mu & \omega \,g \,q\, ZR
    \end{array}
    \right)
\end{align}
In this representation, the diagonal elements of $\hat H_{BO}(R)$, i.e., the diabatic potential energy curves, are parabolas displaced in position, since they are centered in $R_1$ and in $R_2$, and in energy by $\Delta$. The diabatic states are coupled via the off-diagonal elements, which are chosen to have a Gaussian shape centered in $R_3$, with inverse width $a$ and amplitude $b$. Diagonalization of $\hat H_{BO}(R)$ yields the adiabatic potential energy curves, like those shown in Fig.~\ref{fig: scheme} on the left for a particular choice of parameters ($R_1\neq R_2$ and $\Delta =0$), and adiabatic electronic states S$_0$ and S$_1$. The Hamiltonian $\hat V_{P}(q)$ is diagonal and also in this case the potentials are harmonic, centered in zero, with frequency $\omega$ which will be tuned to create on-resonant and off-resonant conditions for the dynamics studies. Finally, the photon-matter Hamiltonian $\hat H_{PM}(q,R)$ is chosen such that the off-diagonal elements feature the transition dipole moment $\mu$ between the two electronic diabatic states, which is constant in space (Condon approximation), whereas the diagonal elements depend only on the nuclear contribution to the dipole $Z R$; the coupling strength is controlled by the parameter $g$. This model Hamiltonian has been implemented in QuantumModelLib.~\cite{QML}

\begin{figure}
    \centering
    \includegraphics[width=1\linewidth]{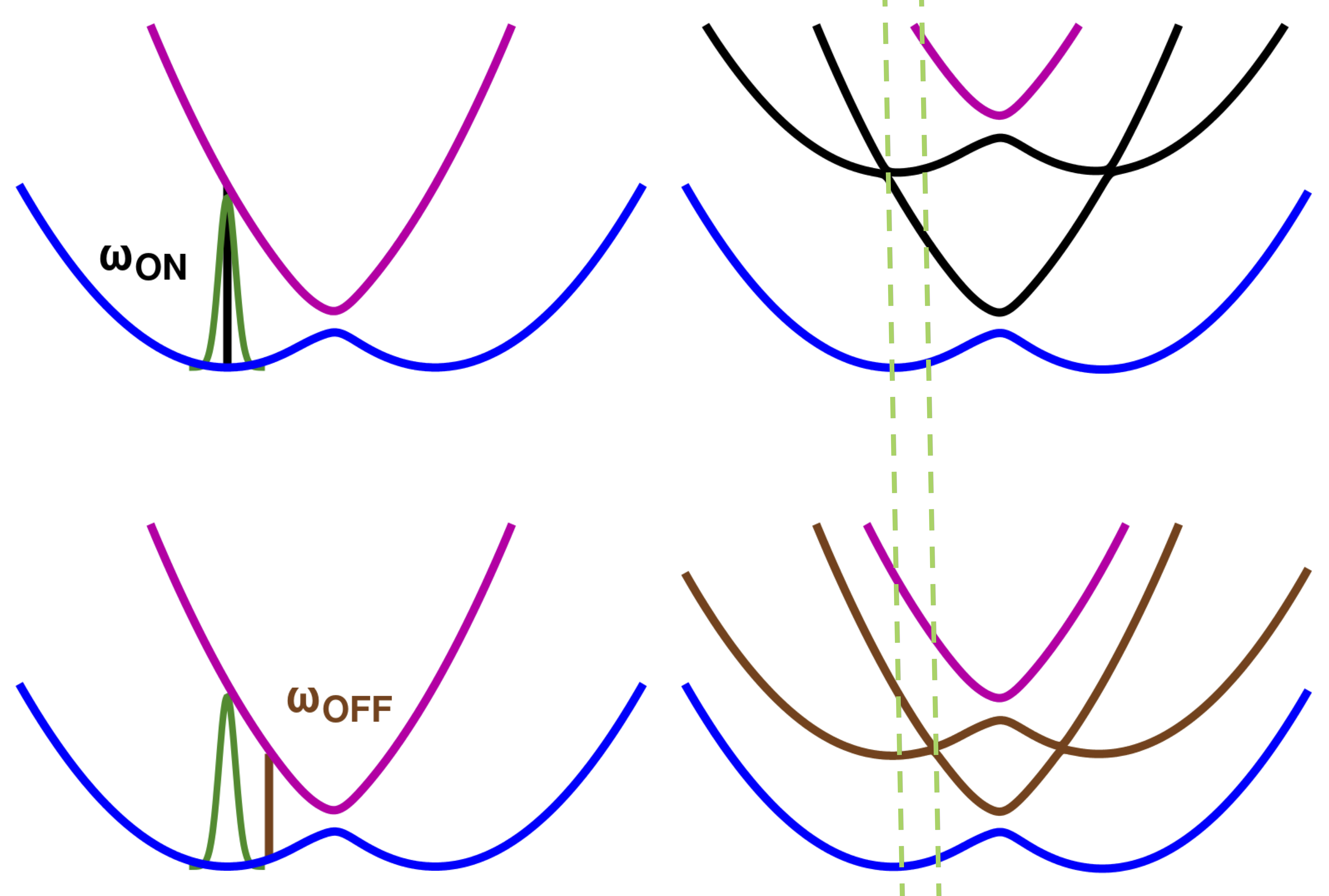}
    \caption{Left panels: bare adiabatic potential energy curves for the model featuring a nonadiabatic process, with the green curves showing the initial nuclear distributions centered at the Franck-Condon point, and the vertical lines identifying the energy gap chosen to tune the frequency of the cavity. Right panels: polaritonic potential energy curves arising in the regime of strong coupling. While the blue and magenta curves are very close in shape to the adiabatic curves on the left, the black curves (top) and the brown curves (bottom) manifest strong hybridization. Due to the different cavity frequencies, the avoided crossings in the black and brown curves arise at different positions, as highlighted by the dashed green vertical lines.}
    \label{fig:highandlowfreq}
\end{figure}

Table~\ref{tab: model parameter} lists the parameters, along with their values, used in the simulations. Two sets of system parameters are defined, yielding respectively a simplified photochemical process with nonadiabatic effects and a photophysical process manifesting Rabi oscillations. The coupling to the cavity mode is controlled by the parameter $g$, which is chosen in all cases to yield strong coupling. Finally, two values of the cavity frequency are identified, which we refer to as ``on-resonance'' and ``off-resonance''. We apply the concepts of ``on-resonance'' and ``off-resonance'' with respect to the Franck-Condon region, which is the region where the nuclear dynamics is initialized (see Fig.~\ref{fig:highandlowfreq} and related discussion).
\begin{table}[h]
\small
  \caption{\ List of the parameters defining the Hamiltonian used in the dynamics studies. Their values are given in atomic units.}
  \label{tab: model parameter}
  \begin{tabular*}{0.48\textwidth}{@{\extracolsep{\fill}}c|c|c}
    \hline
    Parameter & Nonadiabatic process & Rabi oscillations \\
    \hline
    $M$ ($m_e$) & 20000 & 20000 \\
    $a$ (a$_0^{-2}$) & 3.0 & 3.0 \\
    $b$ ($E_h$) & 0.01 & 0.01 \\    
    $k$ ($E_h^2m_e/\hbar^2$) & 0.02 & 0.02 \\
    $R_1, R_2, R_3$ (a$_0$) & 6.0, 2.0, 3.875 & 2.0, 2.0, 3.875\\
    $\Delta$ ($E_h$) & 0 & 0.17 \\
    $\omega$ (on, off) ($E_h$) & 0.17, 0.097 & 0.17, 0.12 \\
    $g$ ($(e\mathrm a_0)^{-1}$ & 0.01 & 0.01 \\
    $\mu$ ($e$a$_0$) & -1.0 &  -1.0 \\
    $Z$ ($e$) & 1.0 & 1.0 \\
    \hline
  \end{tabular*}
\end{table}

In all simulations, the dynamics is initialized in the electronic excited state S$_1$ with a real Gaussian-shaped nuclear wavefunction centered at $R=R_2$~a$_0$ with variance $\sigma_R = \sqrt[4]{(Mk)^{-1}}$. The light field at the initial time is in the ground state, meaning that it is represented as a real Gaussian-shaped wavefunction centered in $q=0$ with variance $\sigma_q=\sqrt[4]{(\omega)^{-1}}$, associated to an average photon number equal to zero. The PEN wavefunction at time $t=0$ is thus
\begin{align}
    \left| \Psi (q,R,t=0)\right\rangle = \chi^{\mathrm{(EL)}}(q,R,t=0) \left| \phi_{\mathrm S_1} (q,R)\right\rangle \\
    = \sqrt[4]{\frac{1}{\pi\sigma_R^2}}e^{-\frac{(R-R_2)^2}{2\sigma_R^2}}\sqrt[4]{\frac{1}{\pi\sigma_q^2}}e^{-\frac{q^2}{2\sigma_q^2}}\left| \phi_{\mathrm S_1} (q,R)\right\rangle
\end{align}
by using the Dirac notation for the electronic discrete representation rather than the position representation. In the photon-number representation, instead, the initial condition reads
\begin{align}
    \left| \Psi (q,R,t=0)\right\rangle = \chi^{\mathrm{(POL)}}(R,t=0) \left| \tilde\phi_{\mathrm S_1} (R)\right\rangle \left|0\right\rangle \\
    = \sqrt[4]{\frac{1}{\pi\sigma_R^2}}e^{-\frac{(R-R_2)^2}{2\sigma_R^2}}\left| \tilde\phi_{\mathrm S_1} (R)\right\rangle\left|0\right\rangle
\end{align}
with the photon-electronic initial condition $| \tilde\phi_{\mathrm S_1} (R)\rangle|0\rangle$ which can be, in turn, expressed as a linear combination of polaritonic states $\lbrace |\varphi_k(R)\rangle\rbrace$. We used the symbol $| \tilde\phi_{\mathrm S_1} (R)\rangle$ to indicate a ``traditional'' adiabatic state, i.e., an eigenstate of $\hat H_{BO}(R)$.

The quantum dynamics simulations are performed employing the electronic perspective using the split-operator technique~\cite{Steiger_JCP1982} to evolve in time the photon-nuclear wavefunctions representing the projections of $\left| \Psi (q,R,t)\right\rangle $ onto the electronic $q,R$-dependent basis. These projections depend on $q$ and $R$, as well as on time, and are represented on grids of 500 points each, defined in the range $q=[-10,10]$~a$_0$ and $R=[-2,10]$~a$_0$. We used a time step of 0.1~a.t.u. From these simulations, the population of the electronic excited state S$_1$ is calculated as well as the average photon number as functions of time, which are used to assess the performance of the trajectory-based schemes.

The quantum-classical simulations with CTMQC, MTE and TSH are performed using the G-CTMQC code~\cite{GCTMQC} which is interfaced with QuantumModelLib. Adopting the electronic perspective, $N_{tr}=10000$ trajectories have been sampled according to the Wigner distribution obtained from $\chi^{\mathrm{(EL)}}(q,R,t=0)$, providing a set of initial conditions for the positions and the momenta of the nuclear motion and of the photonic motion. G-CTMQC propagates the electronic system in the basis defined in Eq.~\eqref{eqn: def EEL} and thus provides direct access to the population of the electronic state S$_1$. The average photon number is, instead, estimated as 
\begin{align}\label{photon-number}
 \langle \hat{a}^\dagger \hat{a} \rangle = \frac{\omega}{2} \langle \hat{q}^2 \rangle + \frac{1}{2\omega} \langle \hat{p}^2 \rangle - \frac{1}{2}
\end{align}
exactly as it is done in the quantum calculation. 
Note that the above expression is only true only up to within zero-th order in $g$, since the physical photon number operator in the dipole gauge is $\hat a^\dagger \hat a+g\hat q(Z\hat R-\hat r) +g^2 (Z\hat R-\hat r)^2/2\omega$.~\cite{Hoffmann2020, Mandal2023} In the used parameters regime, we expect the contributions of order $g$ to the photon number to be small, while the the contributions of order $g^2$ to be strongly suppressed. In addition, since we only study this quantity to compare different theoretical formulations within the same gauge, and for the sake of clarity of the arguments below, we will use the Pauli-Fierz photon number as given in Eq.~\eqref{photon-number}.

Adopting the polaritonic perspective, the initial condition has to be defined identical to the electronic perspective and has to be given in the polaritonic basis, since G-CTMQC performs the propagation in this basis. While the initial sampling along $R$ can be performed as just described above, thus according to the Wigner distribution, the polaritonic initial state has to account for the electronic and for the photonic initialization. 

To clarify the procedure, let us show in Fig.~\ref{fig:highandlowfreq} the polaritonic potential energy curves $E_k^{\mathrm{(POL)}}(R)$ (right) in comparison to adiabatic electronic potential energy curves (left) in the absence of coupling to the cavity. We will first describe the situation using the parameters for the model Hamiltonian mimicking a nonadiabatic process with cavity mode on-resonance ($\omega_{\mathrm{on}}$) in the top panels. In this case, the frequency of the cavity is tuned to be the same as the S$_0$-S$_1$ energy gap at the Franck-Condon point. Since the light field can be in only two photon-number states, a total of four polaritonic states are obtained from Eq.~\eqref{eqn: def EPOL}, which can be identified as the blue and magenta curves as well as the black curves. 
On the one hand, the polaritonic ground state (blue curve) and third-excited state (magenta curve) can be quite clearly identified as the electronic ground state ``dressed'' by the zero-photon state and as the electronic excited state ``dressed'' by the one-photon state. Note that the magenta curve on the right is shifted by one-photon energy but remains otherwise very close in shape to the corresponding curve on the left. On the other hand, the black curves have a strongly-mixed light-matter character, especially at and in the vicinity of the avoided crossings. Similar observations can be presented for the off-resonance case ($\omega_{\mathrm{off}}$) in the bottom panels of Fig.~\ref{fig:highandlowfreq}. However, since the cavity frequency is chosen to be equal to the S$_0$-S$_1$ energy gap away from the Franck-Condon point, the positions of the new avoided crossings formed in the brown polaritonic potential energy curves is different from the on-resonance case, as highlighted by the dashed green vertical lines.

\begin{figure}
    \centering
    \includegraphics[width=1\linewidth]{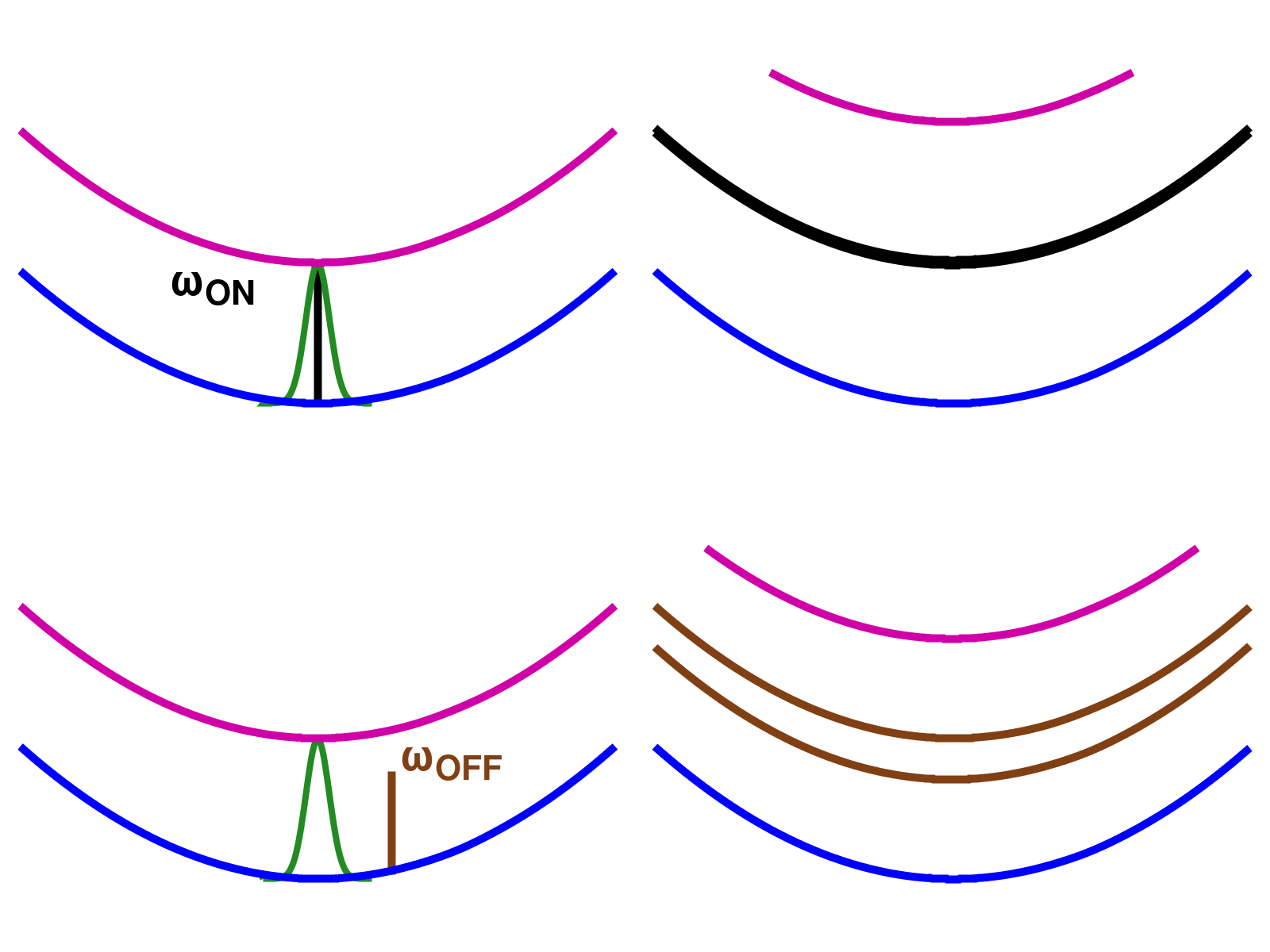}
    \caption{Same as in Fig.~\ref{fig:highandlowfreq} but for the Rabi model.}
    \label{fig:highandlowfreq Rabi}
\end{figure}

Figure~\ref{fig:highandlowfreq Rabi} shows the same kind of potential energy curves as in Fig.~\ref{fig:highandlowfreq} but for the parameters yielding Rabi oscillations. All potential energy curves, both the bare adiabatic curves in the left panels and the polaritonic curves in the right panels, are very close to parabolas which are shifted in energy. In the top right panel, the black curves are almost superimposed do to the resonance condition.

\begin{figure}
    \centering
    \includegraphics[width=1\linewidth]{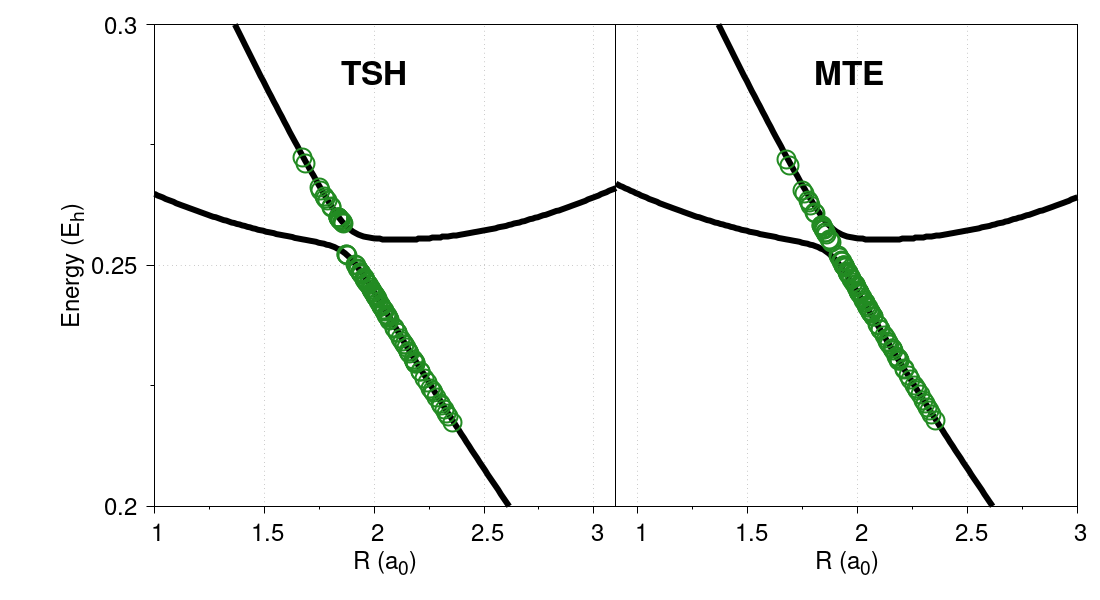}
    \caption{Energy-position distribution of TSH trajectories (left) and MTE/CTMQC trajectories (right) at the initial time in the polaritonic perspective. The trajectories are indicated as green dots.}
        \label{fig:trajonpol_TSHMTE}
\end{figure}

In Fig.~\ref{fig:trajonpol_TSHMTE} it is shown that the avoided crossings in this on-resonance case appear at $R=2$~a$_0$, and the other one at $R=6$~a$_0$. For $R<2$~a$_0$, the lower-energy curve resembles the electronic ground state ``dressed'' by the one-photon state, as it is basically the blue curve shifted by the energy of one photon, whereas the higher-energy curve corresponds to the electronic excited state ``dressed'' by the zero-photon state. These observations hold as well for $R>6$~a$_0$. Instead, in the region between $R=2$~a$_0$ and $R=6$~a$_0$ the opposite is true, namely the  lower-energy curve resembles the electronic excited state ``dressed'' by the zero-photon state and the higher-energy curve resembles the electronic ground state ``dressed'' by the one-photon state. At the avoided crossings, these identifications are not even possible. Therefore, the initialization of the dynamics in the electronic excited state with the light in the ground state, i.e., in the zero-photon state, is achieved by constructing a linear combination of the two polaritonic states associated to the black curves around $R=2$~a$_0$, which is the center of the nuclear distribution at time $t=0$.

Therefore, $| \tilde\phi_{\mathrm S_1} (R)\rangle|0\rangle \simeq C_3^{\mathrm{(POL)}}(R<2~\mathrm{a}_0, t=0)|\varphi_3(R<2~\mathrm{a}_0)\rangle + C_2^{\mathrm{(POL)}}(R>2~\mathrm{a}_0, t=0)|\varphi_2(R>2~\mathrm{a}_0)\rangle $ where $|\varphi_2(R)\rangle$ and $|\varphi_3(R)\rangle$ are the polaritonic states associated to the black potential energy curves in Fig.~\ref{fig:trajonpol_TSHMTE}. The green dots in Fig.~\ref{fig:trajonpol_TSHMTE} show how this information in encoded in the trajectories that will be propagated according to the TSH scheme (left panel) and according to MTE and CTMQC (right panel). Specifically, in the region of the formation of the avoided crossing, where the hybridization of the light excitation and of the molecular excitation is strongest, TSH trajectories have to be associated to one active state, which is either one of the two polaritonic states, whereas MTE and CTMQC trajectories can be associated to coherent superposition of polaritonic states. This difference yields a slight deviation in the energy distribution of the trajectories when comparing the TSH initial condition with the MTE/CTMQC one. Nonetheless, while for every TSH initial trajectory the active state is clearly identified as the state with larger probability $|C_k^{\mathrm{(POL)}}(R, t=0)|^2$ between $k=2$ and $k=3$, the corresponding trajectory can remain associated to a coherent superposition of polaritonic states. While we understand that this choice is not optimal for TSH, the ensuing dynamics shows a very good agreement with reference quantum results. 

In the ``off-resonance'' case, the same observations apply, but the avoided crossings between the brown polaritonic energy curves appear more in the vicinity of the maximum of the blue curve located at approximately $R=4$~a$_0$. In this case, the avoided crossing on the left is slightly outisde of the Franck-Condon region and the initial state corresponding to the electronic excited state with zero photons as a pure polaritonic character, namely $| \tilde\phi_{\mathrm S_1} (R)\rangle|0\rangle \simeq C_3^{\mathrm{(POL)}}(R, t=0)|\varphi_3(R)\rangle$. In this case, all TSH trajectories are associated to the same active state and their energy distribution at time $t=0$ is identical as in MTE and CTMQC.

Note that the quantum-classical dynamics employing the polaritonic perspective yields the contributions of the four polaritonic states as functions of time in terms of $C_k^{\mathrm{(POL)}}(R^{(I)}, t)$ along all trajectories. A change of basis is performed at each time to recover information in the basis $|\tilde\phi_{l} (R)\rangle|n\rangle$ with $l=\mathrm S_0, \mathrm S_1$ and $n=0,1$, which is used to determine the population of the electronic excited state S$_1$ and the average photon number to be compared to the benchmark quantum dynamics. 

In the following sections, we will compare the results of the quantum-classical simulations in the electronic and polaritonic perspectives to the quantum-dynamics results. 

\subsection{Nonadiabatic process}\label{sec: nonadiabatic}

\begin{figure}
    \centering
    \includegraphics[width=\linewidth]{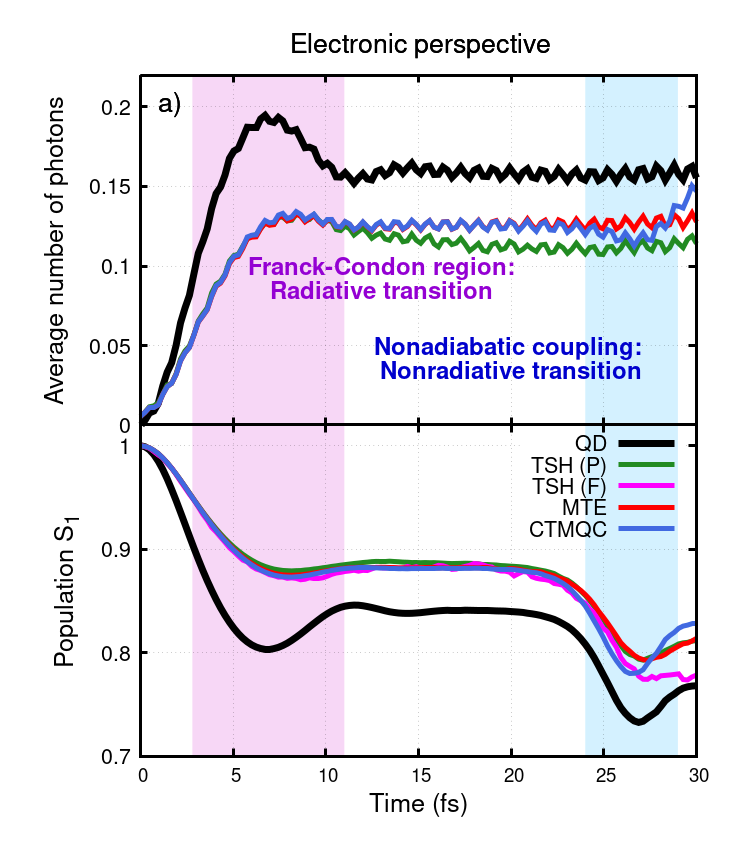}
    \hfill
    \includegraphics[width=\linewidth]{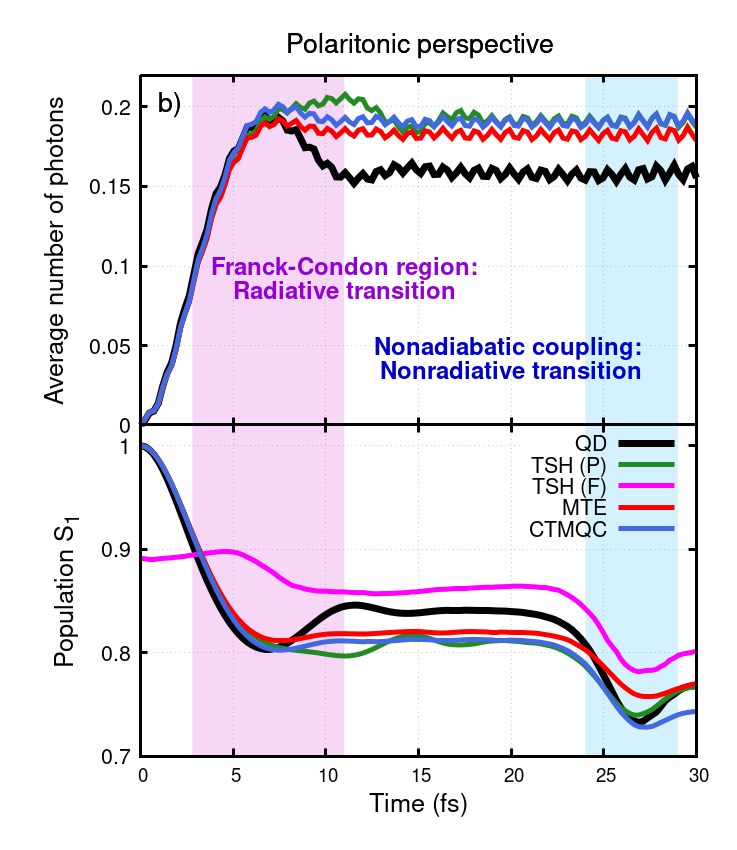}
    \caption{Time trace of the average photon number (top plots) and of the S$_1$ population (bottom plots) for the nonadiabatic model on resonance. Quantum dynamics result are shown in black, TSH in green (P) and magenta (F), MTE in red and CTMQC in blue. a) Electronic perspective. b) Polaritonic perspective. Note that in a) the average photon number with TSH is calculated from the spatial distribution of the trajectories, so it is independent of the way (F or P) the S$_1$ population is estimates even though we used the same color as TSH(P).}
    \label{fig:joined-on}
\end{figure}

The adiabatic potential energy curves in Fig.~\ref{fig:highandlowfreq} (left) represent the illustrative model studied in this section in the absence of strong coupling with the cavity. When a nuclear wavepacket is initialized in the Franck-Condon region associated to the left minimum of the ground state (blue curve) in the excited state, it evolves towards the right until it reaches the avoided crossing between the two electronic states. There, a nonadiabatic event takes place and some population is transferred from the excited state to the ground state while the two portions of the original wavepacket keep moving towards the right (but at different speed after the branching due to the different slopes of the corresponding potentials). When this electron-nuclear system is strongly coupled to a cavity mode, the dynamics is altered since the excited-state population can be transferred to the ground state as consequence of the light-matter coupling, thus yielding radiative decay responsible for photon emission, and as consequence on the nonadiabatic coupling, thus yielding non-radiative decay. The interplay of these events depends on the resonance conditions, which we identify via $\omega_{\mathrm{on}}$ and $\omega_{\mathrm{off}}$ and whose effect is schematically shown in Fig.~\ref{fig:highandlowfreq}.

Figure~\ref{fig:joined-on}a)  shows the results of the quantum-classical dynamics simulated using TSH (green and magenta), MTE (red) and CTMQC (blue) in the electronic perspective. The black lines are the reference quantum results showing the time trace of the average photon number (upper panel) and of the the population of the electronic excited state (lower). Since the cavity is in resonance with the S$_0$-S$_1$ excitation at the Franck-Condon point with $\omega_{\mathrm{on}}$, as soon as the dynamics starts, the excited-state population is transferred to the ground state while at the same time the average photon number increases. This behavior attests to a radiative decay until approximately 10~fs (purple shaded area in the panels of Fig.~\ref{fig:joined-on}). When the wavepacket leaves the Franck-Condon region, the resonance condition is not fulfilled anymore, and the average photon number remains constant until the end of the simulated dynamics. Nonetheless, at later time, i.e., just before 25~fs, a non-radiative decay is observed, since the excited-state population decreases but the average photon number is unaffected. Note that the high-frequency oscillations observed in the average photon number are due to the asymmetry of the potential along the photon displacement direction induced by the coupling to nuclear variable $R$.

The quantum-classical results reproduce fairly well the behavior of the excited-state population and of the average photon number. For MTE and CTMQC, the population of S$_1$ is estimated as
\begin{align}\label{eqn: P S1 for MTE}
  P_{\mathrm S_1}(t) = \frac{1}{N_{tr}} \sum_{I=1}^{N_{tr}} \left|C_{\mathrm S_1}^{(I)}(t)\right|^2  
\end{align}
with a trajectory $\bm X^{(I)}(t)$ identified by the coordinates $R^{(I)}(t),q^{(I)}(t)$ along the nuclear and photonic dimensions. In TSH, one can use Eq.~\eqref{eqn: P S1 for MTE}, which is shown in green in Figure~\ref{fig:joined-on}a) (lower panel) and indicated as TSH(P), as well as
\begin{align}\label{eqn: F S1 for MTE}
  F_{\mathrm S_1}(t) = \frac{N_{\mathrm S_1}(t)}{N_{tr}} 
\end{align}
which is shown in magenta Figure~\ref{fig:joined-on}a) (lower panel) and indicated as TSH(F). The average photon number, instead, is determined as in Eq.~\eqref{photon-number} from the phase-space distribution of the trajectories in the photonic dimension in all simulations, using the expressions
\begin{align}
    \langle \hat{q}^2 \rangle &= \frac{1}{N_{tr}} \sum_{I=1}^{N_{tr}} [q^{(I)}(t)]^2\\
    \langle \hat{p}^2 \rangle &= \frac{1}{N_{tr}} \sum_{I=1}^{N_{tr}} [p^{(I)}(t)]^2
\end{align}
We note that while the profiles are overall quite well reproduced in all cases, the S$_1$ population is always overestimated while the average photon number is underestimated. As discussed in our previous work,~\cite{Agostini_JCP2024_2} this deviation from the exact reference is due to effects related to the small mass associated to the photon displacement coordinate. 

\begin{figure}
    \centering
    \includegraphics[width=\linewidth]{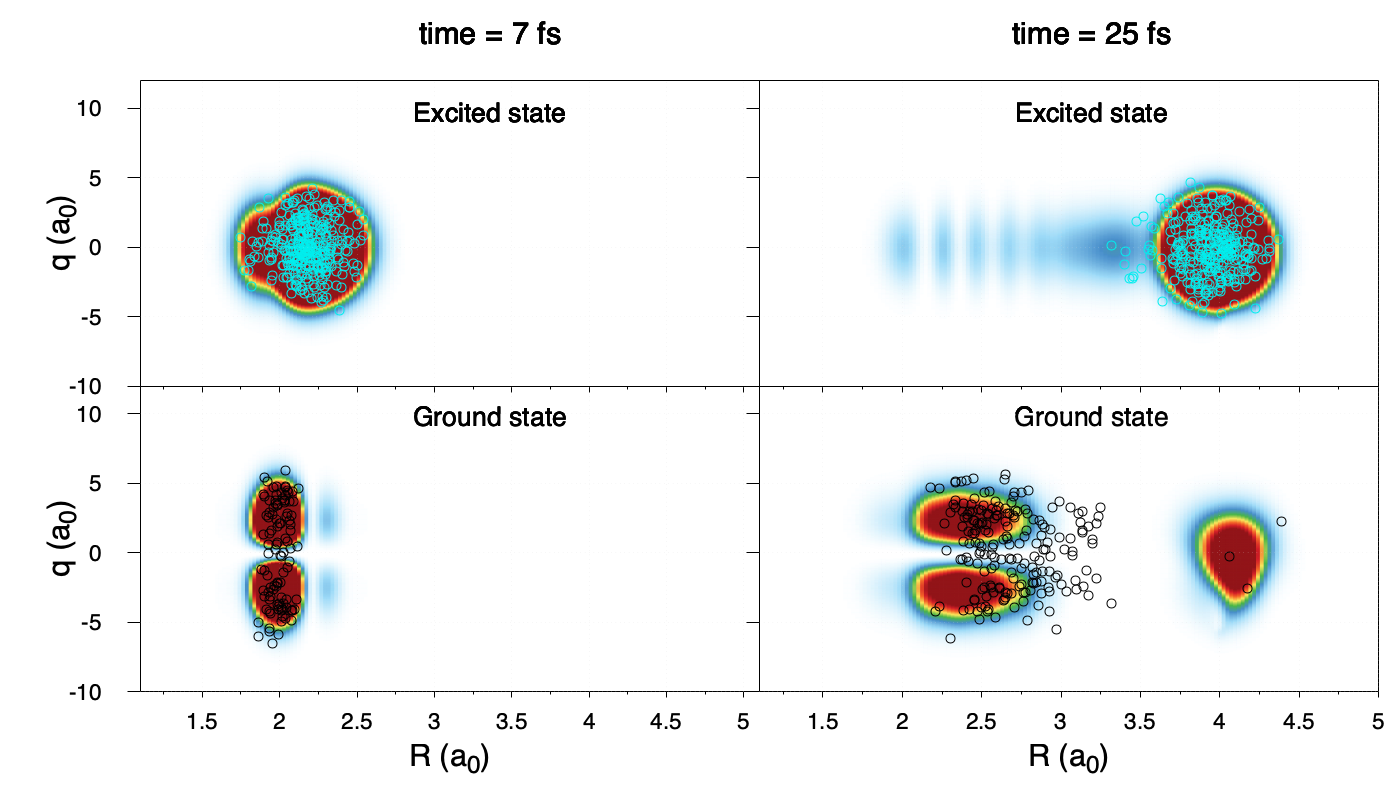}
    \caption{Marginal density in the $q,R$ space at 7~fs (left) and at 25~fs (right), projected onto the electronic excited state (top) and ground state (bottom). CTMQC trajectories are superimposed and distributed according to the dominant electronic state}
    \label{fig:trajonel}
\end{figure}

The classical-like treatment of the photonic degree(s) of freedom is fully equivalent to the treatment of the nuclear dynamics. Nonetheless, for the photons, the ratio with the electronic mass is not as forgiving as for the nuclei. Quantum effects in the trajectories are completely neglected in all quantum-classical schemes, and, in addition, in CTMQC the TDPES is approximated by neglecting the contribution from $\hat U_{coup}$ in Eq.~\eqref{eqn: coup}. While it has been shown~\cite{AgostiniEich_JCP2016} that this is valid approximation in electron-nuclear problems based on the small electron-nuclear mass ratio, it loses validity when light particles are considered (in this particular situation, the electron-photon mass ratio is one). In order to confirm the effect of neglecting $\hat U_{coup}$ in Eq.~\eqref{eqn: coup}, we show in Fig.~\ref{fig:trajonel} the distribution of CTMQC trajectories at two times along the dynamics in comparison to the marginal density, which is a function of $q$ and $R$ in the electronic perspective. The upper panels show the projection of the density onto the excited state while the lower panels report the projections onto the ground state. At 7~fs (left), the density is still localized in the Franck-Condon region, with the portion in the ground state clearly showing a bimodal distribution. At later times, specifically at 25~fs (right), the portion of the density evolving in the excited state undergoes a nonadiabatic event: between $R=4$~a$_0$ and $R=4.5$~a$_0$, the density has contributions both in the excited and in the ground state, but has a unimodal distribution along $q$. At the same time, the portion of the density in the Franck-Condon region in the ground state remains bimodal in the $q$ direction. The interpretation of this dynamics is quite straightforward and interesting, since it is clear that the bimodal distribution in $q$ is associated to the first excited state of the harmonic oscillator representing the photon Hamiltonian and, thus, the one-photon state is populated. However, this event takes place in the Franck-Condon region, while the remaining portion of the density is unimodal and, thus, associated to the zero-photon state. The distribution of CTMQC trajectories in the ground state in the Franck-Condon region is not capable to fully reproduce the spatial splitting along $q$, which causes the average photon number to be underestimated. Previous studies~\cite{Maitra_EPJB2018, Maitra_JCP2022} have confirmed that indeed the term $\hat U_{coup}$ in Eq.~\eqref{eqn: coup} is responsible for the appearance of a large potential barrier capable of producing the splitting of the marginal density in the $q$ direction.
It is worth noting that in Fig.~\ref{fig:trajonel}, CTMQC trajectories have been associated to the ground or to the excited state depending on the values of the corresponding populations, namely $|C_{\mathrm S_0}^{(I)}(t)|^2$ and $|C_{\mathrm S_1}^{(I)}(t)|^2$.

The same dynamics described so far can be simulated with TSH, MTE and CTMQC using the polaritonic perspective, whose results are reported in Figure~\ref{fig:joined-on}b). The initial dynamics is very well captured by the quantum-classical schemes, even though some disagreement is observed in the TSH population time trace (lower panel) between $P_{\mathrm S_1}(t)$ (green) and $F_{\mathrm S_1}(t)$ (magenta). However, this is expected due to the way the initial conditions in TSH have been adapted to accommodate for the fact that an avoided crossing is present in the Franck-Condon region. In all cases, after the initial decrease up to $7-8$~fs, the S$_1$ population is predicted by the quantum-classical methods lower than the reference, meaning that too much population has been transferred to second-excited polaritonic state (of character S$_0$ dressed by the zero-photon state for $R>2$~a$_0$) during the first nonadiabatic event, which is associated to the avoided crossing between the black curves at the Franck-Condon region in Fig.~\ref{fig:highandlowfreq} (right). Consequently, after $7-8$~fs, the average photon number is overestimated by the quantum-classical simulations if compared to the reference. A possible justification for this disagreement is the fact that the trajectories do not enter smoothly in the nonadiabatic region, thus experiencing a slow increase of the magnitude of the nonadiabatic coupling, but are initialized in a region where they are large. Such a hypothesis will be confirmed from the analysis of the results obtained in the off-resonance case. After the plateau between $7-8$~fs and $22-23$~fs, the S$_1$ population undergoes an additional decrease when the trajectories encounter the avoided crossing between the polaritonic ground and first-excited state.

\begin{figure}
    \centering
    \includegraphics[width=\linewidth]{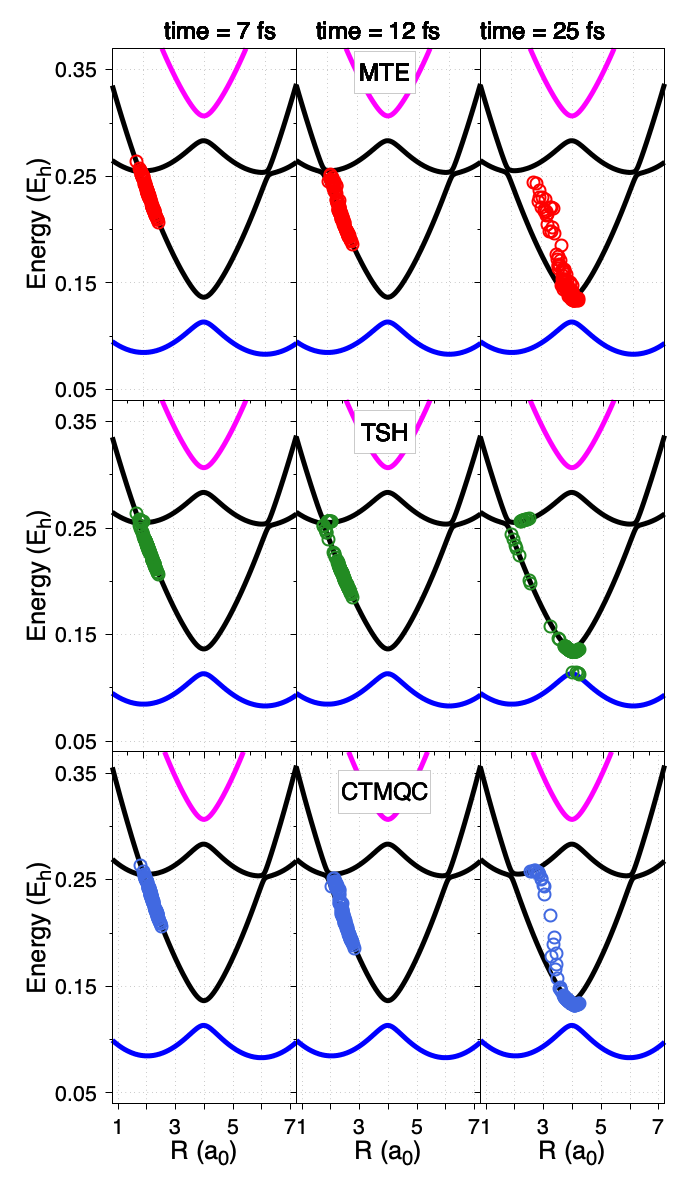}
    \caption{Polaritonic potential energy curves with the distribution of the trajectories superimposed to show their distribution in energy and space at various times, indicated in the figure. MTE trajectories are shown as red dots, TSH trajectories as green dots and CTMQC trajectories as blue dots.}
    \label{fig:semiclassicaltrajectories}
\end{figure}
In the quantum-classical simulations performed in the polaritonic perspective, the electronic population and the average photon number are determined from the coefficients $C_{k}^{(I)}(t)$ propagated alongside the trajectories. However, these coefficients provide information only about the polaritonic states. Therefore, at each time, a basis transformation is performed using the $R$-dependent transformation matrix calculated by QuantumModelLib allowing us to adopt the $|\tilde\phi_k(R)\rangle|n\rangle$ representation, with $k=\mathrm S_0,\mathrm S_1$ and $n=0,1$. In this way, the electronic populations are obtained by summing over all possible number states and the average photon number if obtained by summing over all possible electronic states.

\begin{figure}
    \centering
    \includegraphics[width=\linewidth]{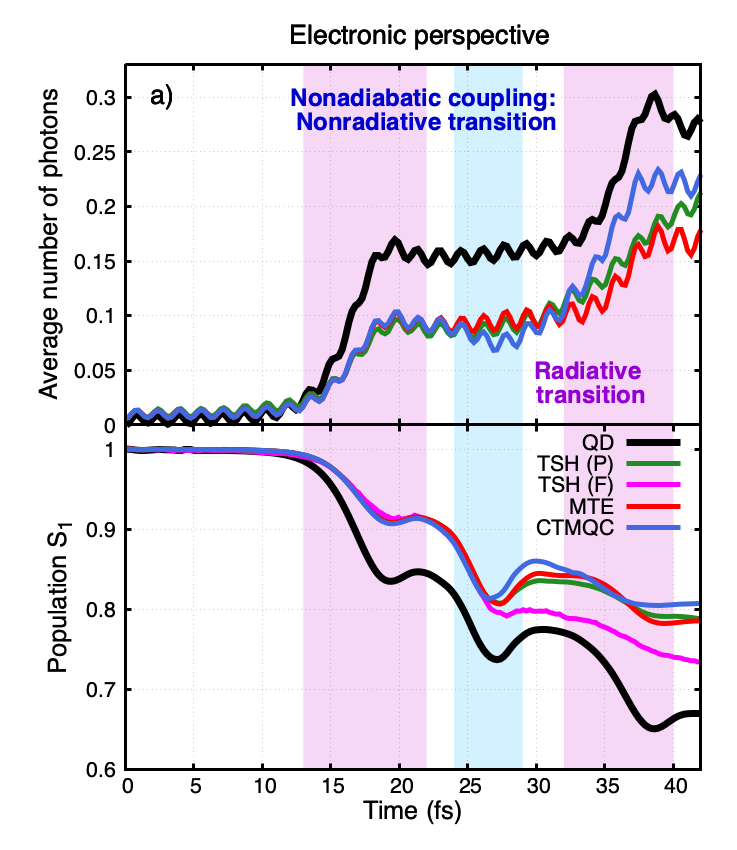}
    \hfill
    \includegraphics[width=\linewidth]{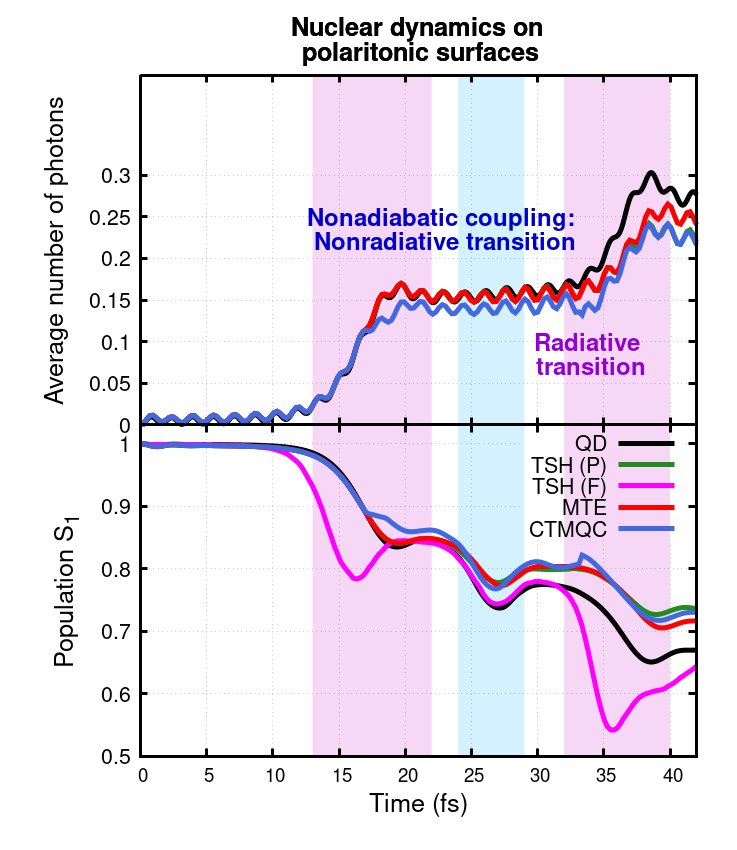}
    \caption{Same as in Fig.~\ref{fig:joined-on} but for the case of the nonadiabatic process off-resonance.}
    \label{fig:joined-off}
\end{figure}
Note that since $F_{\mathrm S_1}(t)$ of Eq.~\eqref{eqn: F S1 for MTE} does not give access to a proper vector of coefficients, the average photon number cannot be estimated correctly using this information.

The observations on the dynamics reported above are clearly drawn by analyzing Fig.~\ref{fig:semiclassicaltrajectories}, where the four polaritonic potential energy curves are shown as reference for the distributions of the trajectories in energy and in $R$ space at different times calculated by MTE (top panels), TSH (middle panels) and CTMQC (bottom panels). MTE trajectories (red dots in Fig.~\ref{fig:semiclassicaltrajectories}) show some delocalization both in energy and in space at all times, but their distribution remains more compact than TSH and CTMQC in both directions, which is expected due to the mean-field character of the approximations. TSH trajectories (green dots in Fig.~\ref{fig:semiclassicaltrajectories}), instead, move only on the (active) polaritonic energies, and, for instance at the last time step shown in the figure, they are distributed in the ground state (blue curve associated to S$_0$ with zero photons), in the first-excited state (lower black curve associated to S$_1$ with zero photons) and in the second-excited state (higher black curve associated to S$_0$ with one photon for $R>2$~a$_0$). CTMQC trajectories (blue dots in Fig.~\ref{fig:semiclassicaltrajectories}) follow an approximate TDPES, and in some portions of $R$ they can be associated to the polaritonic energies, similarly to TSH. A direct relation can be drawn between Fig.~\ref{fig:semiclassicaltrajectories} and Fig.~\ref{fig:trajonel}: the trajectories found at around $R=4$~a$_0$ in Fig.~\ref{fig:trajonel} are partially in the excited state and partially in the ground state, but show a strong unimodal distribution, and in fact in Fig.~\ref{fig:semiclassicaltrajectories} they are associated to the polaritonic potential energy curves carrying a zero-photon character; the trajectories found in Fig.~\ref{fig:trajonel} between $R=2.5$~a$_0$ and $R=3.5$~a$_0$ are mainly associated to the ground state with a bimodal distribution, and consequently in Fig.~\ref{fig:semiclassicaltrajectories} are associated to the electronic ground state dressed by the energy of one photon. It is worth noting that in all quantum-classical calculations, the third-excited polaritonic state remains not populated as in the reference quantum dynamics.


Figure~\ref{fig:joined-off} show results analogous to Fig.~\ref{fig:joined-on} but setting the cavity mode ``off-resonance'', namely choosing $\omega_{\mathrm{off}}$ smaller than $\omega_{\mathrm{on}}$ such that it is in resonance with the S$_0$-S$_1$ excitation outside of the Franck-Condon region. This is schematically depicted in Fig.~\ref{fig:highandlowfreq} in the bottom panels.

Employing the electronic perspective in Figure~\ref{fig:joined-off}a), the average photon number remains underestimated all along the simulated dynamics by all quantum-classical methods, with a consequent prediction of the S$_1$ population that is overestimated, with respect to the reference quantum results. This disagreement can be ascribed to the use of the classical approximation for the evolution of very light degrees of freedom, i.e., the photons, as done previously. The overall dynamics is, however, slightly different compared to the previous study, since over the duration of the simulated dynamics, the wavepackets/trajectories encounter twice a region of radiative transitions, identified by the purple shaded areas in Fig.~\ref{fig:joined-off}, before and after the region of non-radiative transition, identified by the cyan shaded area in Fig.~\ref{fig:joined-off}. The time trace of the S$_1$ population in Fig.~\ref{fig:joined-off}a) (lower panel) shows good agreement among all quantum-classical methods, with a deviation of the magenta line corresponding to $F_{\mathrm S_1}(t)$ in TSH just after 25~fs. It is not completely unexpected that $F_{\mathrm S_1}(t)\neq P_{\mathrm S_1}(t)$ especially since this happens during and after the crossing of a region of weak nonadiabatic coupling between the electronic ground and first-excited states. Nonetheless, the profile of the remaining S$_1$ curves agrees qualitatively with the reference.

A better agreement between the quantum-classical simulations and the reference is observed in Fig.~\ref{fig:joined-off}b), which reports the results obtained by employing the polaritonic perspective. In this case, since the formation of the avoided crossings due to the strong coupling with the cavity mode are outside of the Franck-Condon region, as schematically depicted in Fig.~\ref{fig:highandlowfreq} (bottom), the initialization of all simulations is done in the same way, since all trajectories are associated to the second-excited polaritonic state which has a S$_1$ character dressed by the zero-photon state. While the average photon number is predicted by the quantum-classical methods in good agreement with the quantum dynamics, the S$_1$ population starts deviating from the reference during the passage via the avoided crossing between the polaritonic ground and first-excited states. It is also there that we observe a large disagreement between $F_{\mathrm S_1}(t)$ and $P_{\mathrm S_1}(t)$ during the TSH dynamics. 

\subsection{Rabi oscillations}
Tuning the parameters of the model Hamiltonian defined in Section~\ref{sec: results}, we construct here an illustrative example of Rabi oscillations between the molecular excitation and the light field. The electronic adiabatic potential energy curves are almost parallel parabolas displaced in energy (but not in space), as shown in Fig.~\ref{fig:highandlowfreq Rabi}. In the absence of strong coupling between the system and the cavity mode, the initial condition is represented by the vibrational ground state of the electronic ground state instantaneously promoted to the excited state. In these conditions, the photoexcited wavepacket is stationary, and no non-trivial dynamics is observed.

\begin{figure}
    \centering
    \includegraphics[width=\linewidth]{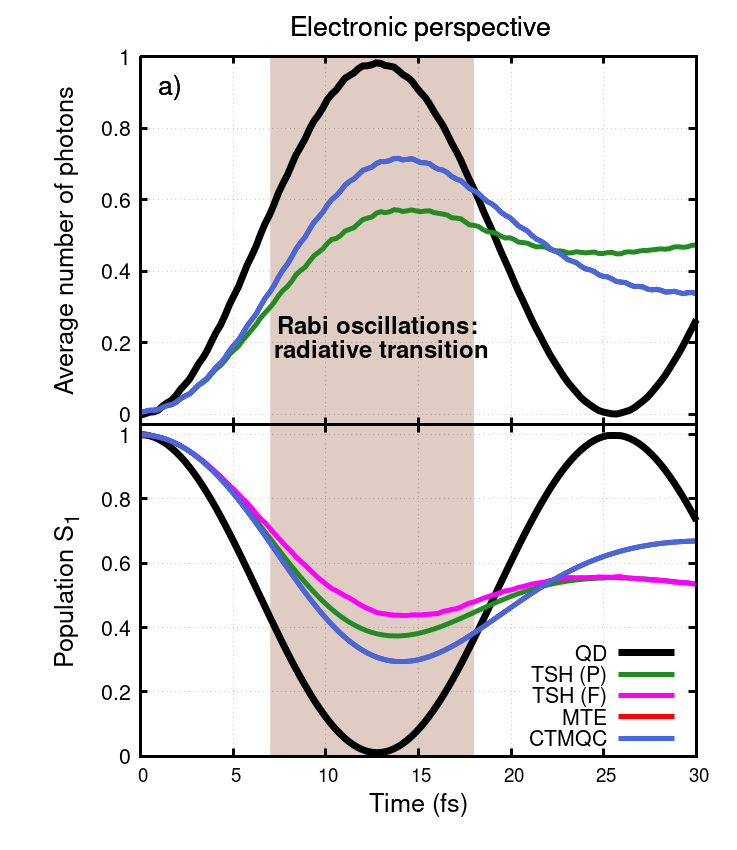}
    \hfill
    \includegraphics[width=\linewidth]{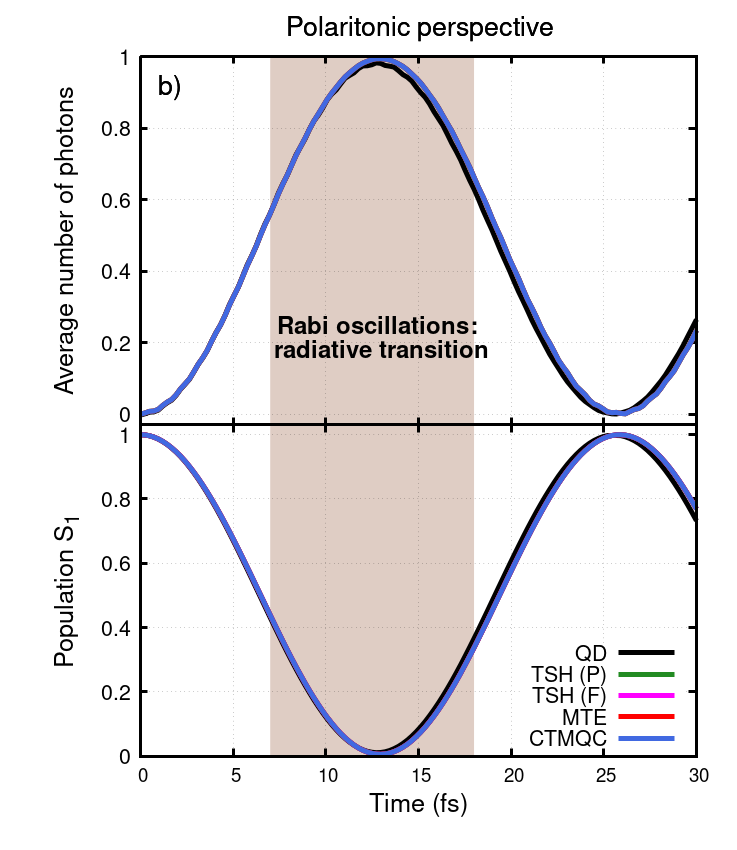}
    \caption{Time trace of the average photon number (top plots) and of the S$_1$ population (bottom plots) for the Rabi oscillation model on resonance. Quantum dynamics result are shown in black, TSH in green and magenta, MTE in red and CTMQC in blue. a) Electronic perspective. b) Polaritonic perspective.}
    \label{fig:Bmodeljoinedon}
\end{figure}

When the system is, instead, strongly coupled to the cavity mode of frequency $\omega_{\mathrm{on}}$, i.e., on-resonance with the S$_0$-S$_1$ excitation, Rabi oscillations emerge after the initial photoexcitation. Specifically, the system transfers energy to the light field, a process that manifests itself with the emission of photon(s). In turn, the light energy is then reabsorbed by the system, and this energy exchange repeats itself periodically over time.

Quantum dynamics simulations clearly show Rabi oscillations between the molecular and the light excitation, which can be identified from the time trace of the population of the electronic excited state and of the average photon number in Fig.~\ref{fig:Bmodeljoinedon}a), lower and upper panel, respectively. In this on-resonance condition, the electronic excited state, which is fully populated at time $t=0$, loses completely its population after a certain time, which corresponds to the time the average number of photons reaches unity. The emitted photon is then reabsorbed by the system and the population of S$_1$ reaches the value one again, within a full Rabi period (which we estimate to about 27~fs). As observed in previous work,~\cite{Agostini_JCP2024_2} the quantum dynamics in the electronic perspective is quite simple along the nuclear direction, as the marginal density remains basically unchanged along $R$, whereas along the photon displacement direction the density oscillates between a unimodal distribution and a bimodal distribution, centered in $q=0$, as the light field switches from the zero-photon state to the one-photon state. The quantum-classical simulations of this process in the electronic perspective capture the oscillations but they do not quantitative agree with the reference. This disagreement is, once again, due to the classical approximation which cannot correctly capture the dynamics along the photon displacement direction. Note that CTMQC and MTE results are superimposed to each other all along the simulated dynamics, and show a slightly better agreement with the reference than TSH. This hypothesis on the reason why the quantum-classical results in Fig.~\ref{fig:Bmodeljoinedon}a) deviate from the quantum results is indeed confirmed when the same process is simulated with TSH, MTE and CTMQC adopting the polaritonic perspective. In this case, all quantum-classical results agree with each other and with reference, as shown in both panels of Fig.~\ref{fig:Bmodeljoinedon}b).

\begin{figure}[!t]
    \centering
    \includegraphics[width=\linewidth]{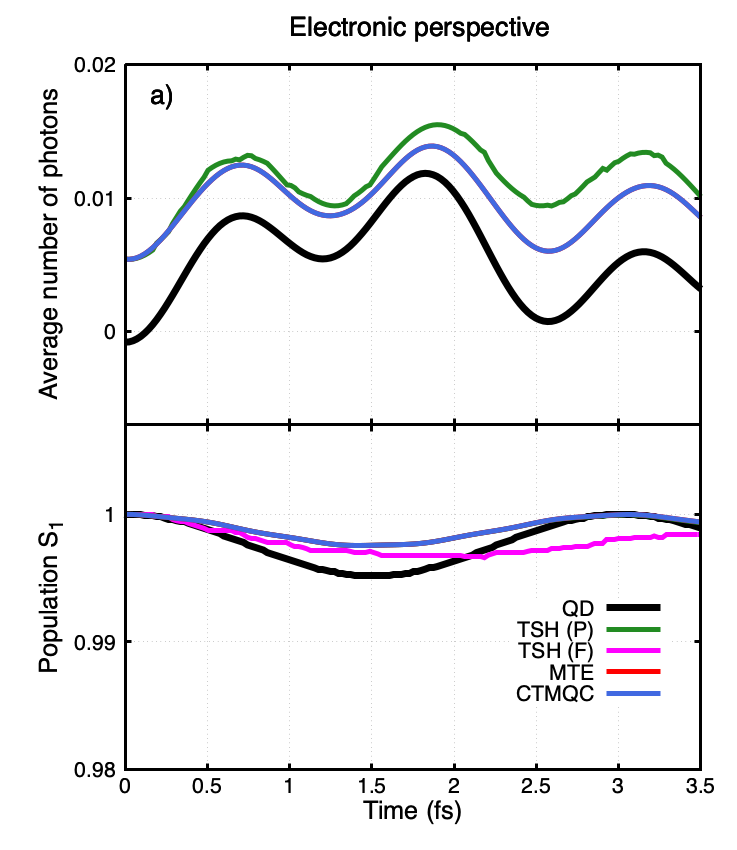}
    \hfill
    \includegraphics[width=\linewidth]{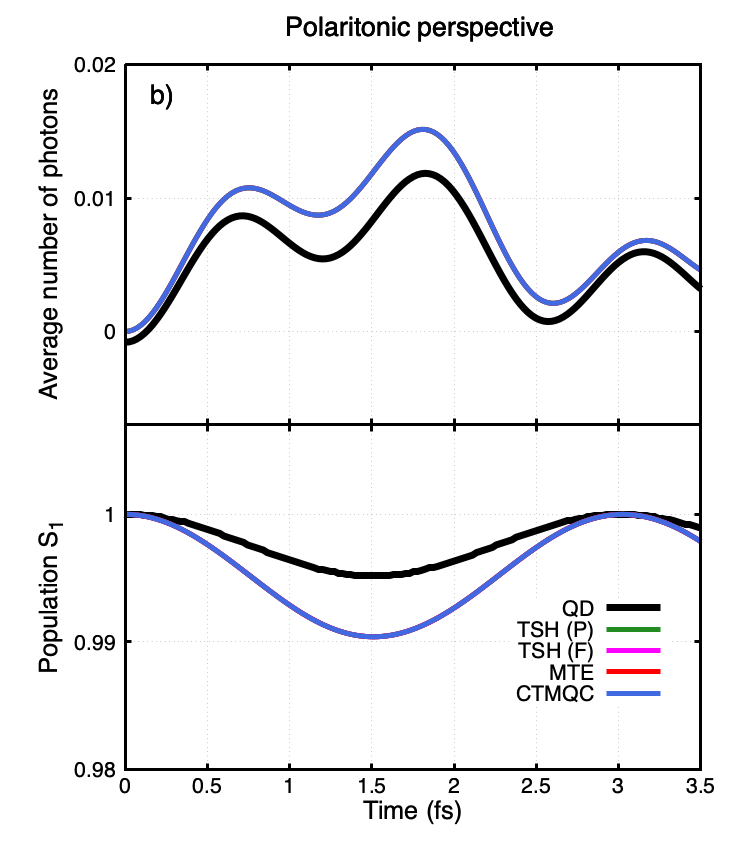}
    \caption{Same as in Fig.~\ref{fig:Bmodeljoinedon} but for the case of the Rabi oscillations off-resonance.}
    \label{fig:Bmodeljoinedoff}
\end{figure}

Similarly to the analysis reported on the nonadiabatic process in Section~\ref{sec: nonadiabatic}, we investigate also in this case the dynamics off-resonance. The results of the quantum dynamics simulations and of the quantum-classical simulations are reported in Fig.~\ref{fig:Bmodeljoinedoff}a) using the electronic perspective and in Fig.~\ref{fig:Bmodeljoinedoff}b) using the polaritonic perspective. Quantum dynamics simulations show indeed oscillations in the S$_1$ population and in the average photon number, but their amplitude as well as their period is strongly reduced if compared to the on-resonance condition. This is an expected result. Specifically, from the oscillation period of $T_{\mathrm{on}}= 27$~fs $= 1116.22$~a.t.u. obtained in Fig.~\ref{fig:Bmodeljoinedon}, we determine the Rabi frequency of $\Omega_{\mathrm{on}}= \frac{2\pi}{T_{\mathrm{on}}} \approx 0.00563~E_h$. In the off-resonance case, the frequency of the Rabi oscillations increases as $\Omega_{\mathrm{off}}= \sqrt{\delta^2 + \Omega^2_{\mathrm{on}}}$ with $\delta = \omega_{\mathrm{on}}-\omega_{\mathrm{off}} = 0.05~E_h$ the difference between the energy gap of the system -- which matches $\omega_{\mathrm{on}}$ -- and the energy of the light field. In the off-resonance case, then, $\Omega_{\mathrm{off}}=0.05030~E_h$ and $T_{\mathrm{off}}= 3.02$~fs, as can be observed in the lower plot of Fig.~\ref{fig:Bmodeljoinedoff}a) showing the time trace of the S$_1$ population. Note that the average photon number shows additional oscillations of higher frequency, exactly as in Fig.~\ref{fig:joined-on}.

The comparison between quantum-classical results and the reference, employing both the electronic and the polaritonic perspectives, are quite satisfying in the off-resonance conditions, as they all agree with each other and they do not deviate strongly from the reference.

\section{Conclusions}\label{sec: conclusions}
We reported an in-depth analysis of dynamics in the presence of strong coupling between light and matter, focusing specifically on different theoretical formulations of the problem employing the exact factorization of the photon-electron-nuclear wavefunction.

First, we proposed two strategies to factor the photon-electron-nuclear wavefunction in terms of a marginal amplitude and a conditional amplitude. Depending if the marginal amplitude is a function of the photon-nuclear coordinates or of the nuclear coordinates, we defined the electronic perspective or the polaritonic perspective, respectively. While the two strategies can be formulated in a very similar way, the ensuing interpretations of the physical processes at hand are qualitatively different. In addition, when developing approximations to introduce a quantum-classical treatment of the photon-electron-nuclear problem which can be adapted to nonadiabatic molecular dynamics, the performance of the classical treatment of the marginal degrees of freedom need to be carefully assessed.

Second, we employed the electronic and the polaritonic perspectives to simulate the dynamics in illustrative model Hamiltonians featuring a nonadiabatic process and Rabi oscillations in the presence of strong light-matter coupling. Also, we induced different coupling conditions by tuning the frequency of the cavity mode on-resonance and off-resonance with respect to the S$_0$-S$_1$ excitation in the Franck-Condon region.

Third, the simulations were performed employing the CTMQC algorithm derived from the exact factorization by introducing the quantum-classical treatment of the dynamics, but connections were made to other widely used schemes, namely TSH and MTE.

The comparison of the two perspectives in the quantum-classical simulation of different processes in different conditions led us to conclude that the polaritonic perspective appears more accurate since it does not invoke the classical treatment of the dynamics of the photonic degrees of freedom, which might break down due to the very small mass associated to the photon displacement coordinate. Nonetheless, the interpretation of the dynamics, for instance when monitoring the time evolution of the population of the electronic states, seems more natural when adopting the electronic perspective, since it does not require changes of representation along the dynamics to switch from the polaritonic basis to the electronic adiabatic basis (dressed by the photon number). Indeed, the polaritonic basis seems better suited for the quantum-classical simulations in the polaritonic perspective since all couplings among the states become nonadiabatic-like.

Despite the discussed limitations of the electronic perspective combined with quantum-classical schemes, we believe that it still represents an interesting avenue for the simulation of photon-electron-nuclear dynamics especially since it stimulates further theoretical developments, aiming to refine the underlying classical treatment of the photon degrees of freedom. Furthermore, we expect the computational cost for the treatment of the photon-electron-nuclear dynamics in the electronic perspective to be comparable to usual electron-nuclear problems. To investigate this point, studies on the application of the electronic perspective are currently ongoing, specifically focusing on how to account for the dependence of the electronic states on the photon displacement in standard electronic-strucure theories.

\section*{Acknowledgements}
This work was supported by the French Agence Nationale de la Recherche via the projects Q-DeLight (Grant No. ANR-20-CE29-0014) and STROM (Grant No. ANR-23-ERCC-0002).

\appendix
\section{CTMQC equations}\label{app: algorithms}
This appendix reports the expressions of the mean-field (mf) and coupled-trajectory (ct) terms in Eqs.~\eqref{eqn: F in CTMQC} and~\eqref{eqn: Cdot in CTMQC} defining the CTMQC algorithm.

The mean-field force acting on the particle indexed by $\Gamma$ at time $t$ is
\begin{widetext}
\begin{align}
\mathbf F^{(I)}_{\Gamma,\mathrm{mf}}(t) = -\sum_k |C_k^{(I)}(t)|^2 \boldsymbol\nabla_\Gamma E_{k}^{(I)}-\sum_{k,l} \bar{C}_l^{(I)}(t) C_k^{(I)}(t)(E_{k}^{(I)}-E_{l}^{(I)}) \bm{d}_{kl,\Gamma}^{(I)} 
\end{align}
\end{widetext}
where $E_{k}^{(I)}$ is the $k$-th energy eigenvalue of the electronic~\eqref{eqn: BH in electronic} or polaritonic~\eqref{eqn: BH in polaritonic} stationary Schr\"odinger equation, calculated at the position $\bm X^{(I)}(t)$, and $\bm{d}_{kl,\Gamma}^{(I)}$ is the nonadiabatic coupling vector between the electronic~\eqref{eqn: BH in electronic} $\langle\phi_k(\bm X^{(I)}(t))|\boldsymbol\nabla_\Gamma\phi_l(\bm X^{(I)}(t)) \rangle_{\bm x}$ or polaritonic~\eqref{eqn: BH in polaritonic} $\langle\varphi_k(\bm X^{(I)}(t))|\boldsymbol\nabla_\Gamma\varphi_l(\bm X^{(I)}(t)) \rangle_{\bm x}$ eigenstates. The symbols $\bar{C}_l^{(I)}(t)$ stands for the complex conjugated of $C_l^{(I)}(t)$.

The coupled-trajectory force is
\begin{widetext}
\begin{align}
\mathbf F^{(I)}_{\Gamma,\mathrm{ct}}(t) = \sum_{k,l} \sum_{\Gamma'}\left(\mathcal{P}_{\Gamma'}^{(I)(t)} \cdot \bm{f}_{k,\Gamma'}^{(I)} \right)\left(\bm{f}_{k,\Gamma}^{(I)} -\bm{f}_{l,\Gamma}^{(I)}\right)|C_k^{(I)}(t)|^2|C_l^{(I)}(t)|^2
\end{align}
\end{widetext}
where $\mathcal{P}_\Gamma^{(I)}(t)=-\boldsymbol\nabla_\Gamma|\chi^{(I)}(t)|^2/(2|\chi^{(I)}(t)|^2)$ is the quantum momentum, and $\bm{f}_{k,\Gamma}^{(I)}=\int_0^t(-\boldsymbol\nabla_\Gamma E_k^{(I)}) d\tau$ being the force associated to the state $k$ accumulated along the trajectory $I$. As in the usual formulation,~\cite{Agostini_JCTC2020_1, Agostini_JCTC2021} the marginal density $|\chi^{(I)}(t)|^2$ in CTMQC is determined as the sum of frozen Gaussians centered at the positions of the trajectories, requiring that, at the end of each dynamics step, the trajectories share information about their positions to compute the quantum momentum. Note that this term is not present in the MTE algorithm.

The mean-field term in the evolution equation of the coefficients representing the conditional amplitude in the electronic~\eqref{eqn: BH in electronic} or polaritonic~\eqref{eqn: BH in polaritonic} basis is
\begin{align}
\dot C_{k,\mathrm{mf}}^{(I)}(t)=-i E_k^{(I)}C_k^{(I)}(t) - \sum_\Gamma \dot{\bm X}_\Gamma^{(I)}(t) \cdot \sum_l \bm{d}_{kl,\Gamma}^{(I)} C_l^{(I)}(t)
\end{align}
which is common to TSH, MTE and CTMQC. Here, we used the symbol $\dot{\bm X}_\Gamma^{(I)}(t)$ indicating the velocity of the trajectory $I$.

The coupled-trajectory term in the coefficients evolution, proper only of CTMQC, is
\begin{align}
\dot C_{k,\mathrm{ct}}^{(I)}(t)= \sum_\Gamma\frac{\mathcal{P}_\Gamma^{(I)}(t)}{M_\Gamma} \cdot \left(\bm{f}_{k,\Gamma}^{(I)} - \sum_l |C_l^{(I)}(t)|^2\bm{f}_{l,\Gamma}^{(I)}\right)C_k^{(I)}(t)
\end{align}

\balance

\newpage


\begin{thebibliography}{100}

\bibitem{Hutchison2012}
J.~A. Hutchison, T.~Schwartz, C.~Genet, E.~Devaux, and T.~W. Ebbesen,
  ``Modifying chemical landscapes by coupling to vacuum fields,'' {\em Angew.
  Chem. Int. Ed.}, vol.~51, p.~1592?1596, Jan. 2012.

\bibitem{Ebbesen_ACR2016}
T.~W. Ebbesen, ``Hybrid light-matter states in a molecular and material science
  perspective,'' {\em Acc. Chem. Res.}, vol.~49, p.~2403, 2016.

\bibitem{Rashidi2025}
K.~Rashidi, E.~Michail, B.~Salcido-Santacruz, Y.~Paudel, V.~M. Menon, and M.~Y.
  Sfeir, ``Efficient and tunable photochemical charge transfer via long-lived
  bloch surface wave polaritons,'' {\em Nat. Nanotechnol.}, vol.~20,
  p.~1618?1624, Aug. 2025.

\bibitem{Garg2025}
P.~Garg, J.~Singh, A.~K. Gaur, S.~Venkataramani, C.~Sch\"{a}fer, and J.~George,
  ``Unveiling the role of dark states in dynamic control of azopyrrole
  photoisomerization by light-matter interaction,'' {\em Commun. Chem.},
  vol.~8, p.~192, July 2025.

\bibitem{Fregoni2018}
J.~Fregoni, G.~Granucci, E.~Coccia, M.~Persico, and S.~Corni, ``Manipulating
  azobenzene photoisomerization through strong light-molecule coupling,'' {\em
  Nat. Commun}, vol.~9, no.~1, p.~4688, 2018.

\bibitem{Groenhof2024}
I.~Sokolovskii and G.~Groenhof, ``Photochemical initiation of
  polariton-mediated exciton propagation,'' {\em Nanophotonics}, vol.~13,
  no.~14, pp.~2687--2694, 2024.

\bibitem{Groenhof2022}
R.~H. Tichauer, D.~Morozov, I.~Sokolovskii, J.~J. Toppari, and G.~Groenhof,
  ``Identifying vibrations that control non-adiabatic relaxation of polaritons
  in strongly coupled molecule-cavity systems,'' {\em J. Phys. Chem. Lett.},
  vol.~13, no.~27, pp.~6259--6267, 2022.

\bibitem{Feist2018}
J.~Feist, J.~Galego, and F.~J. Garcia-Vidal, ``Polaritonic chemistry with
  organic molecules,'' {\em ACS Photonics}, vol.~5, no.~1, pp.~205--216, 2018.

\bibitem{Feist2022}
J.~Fregoni, F.~J. Garcia-Vidal, and J.~Feist, ``Theoretical challenges in
  polaritonic chemistry,'' {\em ACS Photonics}, vol.~9, no.~4, pp.~1096--1107,
  2022.

\bibitem{Subotnik2022-2}
T.~E. Li, B.~Cui, J.~E. Subotnik, and A.~Nitzan, ``Molecular polaritonics:
  Chemical dynamics under strong light-matter coupling,'' {\em Annu. Rev. Phys.
  Chem.}, vol.~73, pp.~43--71, 2022.

\bibitem{Mandal2023}
A.~Mandal, M.~A. Taylor, B.~M. Weight, E.~R. Koessler, X.~Li, and P.~Huo,
  ``Theoretical advances in polariton chemistry and molecular cavity quantum
  electrodynamics,'' {\em Chem. Rev.}, vol.~123, no.~16, pp.~9786--9879, 2023.

\bibitem{Flick2017_1}
J.~Flick, M.~Ruggenthaler, H.~Appel, and A.~Rubio, ``Atoms and molecules in
  cavities, from weak to strong coupling in quantum-electrodynamics (qed)
  chemistry,'' {\em PNAS}, vol.~114, pp.~3026--3034, Mar. 2017.

\bibitem{Ribeiro2018}
R.~F. Ribeiro, L.~A. Mart{\'i}nez-Mart{\'i}nez, M.~Du,
  J.~Campos-Gonzalez-Angulo, and J.~Yuen-Zhou, ``Polariton chemistry:
  controlling molecular dynamics with optical cavities,'' {\em Chem. Sci.},
  vol.~9, no.~30, pp.~6325--6339, 2018.

\bibitem{Yuen-Zhou2026-tt}
J.~Yuen-Zhou, N.~C. Giebink, and R.~F. Ribeiro, eds., {\em Polariton
  chemistry}.
\newblock Nashville, TN: John Wiley {\&} Sons, Jan. 2026.

\bibitem{Ruggenthaler2023}
M.~Ruggenthaler, D.~Sidler, and A.~Rubio, ``Understanding polaritonic chemistry
  from ab initio quantum electrodynamics,'' {\em Chem. Rev.}, vol.~123, no.~19,
  pp.~11191--11229, 2023.

\bibitem{Taylor2025}
M.~A.~D. Taylor, A.~Mandal, and P.~Huo, ``Light?matter interaction hamiltonians
  in cavity quantum electrodynamics,'' {\em Chem. Phys. Rev.}, vol.~6,
  p.~011305, Feb. 2025.

\bibitem{Rubio2018}
M.~Ruggenthaler, N.~Tancogne-Dejean, J.~Flick, H.~Appel, and A.~Rubio, ``From a
  quantum-electrodynamical light-matter description to novel spectroscopies,''
  {\em Nat. Rev. Chem.}, vol.~2, p.~0118, 2018.

\bibitem{Mukamel2023}
B.~Gu, Y.~Gu, V.~Y. Chernyak, and S.~Mukamel, ``Cavity control of molecular
  spectroscopy and photophysics,'' {\em Accounts of Chemical Research},
  vol.~56, no.~20, pp.~2753--2762, 2023.

\bibitem{Kowalewski2016}
M.~Kowalewski, K.~Bennett, and S.~Mukamel, ``Cavity femtochemistry:
  Manipulating nonadiabatic dynamics at avoided crossings,'' {\em J. Phys.
  Chem. Lett.}, vol.~7, no.~11, pp.~2050--2054, 2016.

\bibitem{Angelico2023}
S.~Angelico, T.~S. Haugland, E.~Ronca, and H.~Koch, ``Coupled cluster cavity
  {B}orn-{O}ppenheimer approximation for electronic strong coupling,'' {\em J.
  Chem. Phys.}, vol.~159, no.~21, p.~214112, 2023.

\bibitem{Riso2022}
R.~R. Riso, T.~S. Haugland, E.~Ronca, and H.~Koch, ``Molecular orbital theory
  in cavity {QED} environments,'' {\em Nat. Commun.}, vol.~13, p.~1368, 2022.

\bibitem{Haugland2020}
T.~S. Haugland, E.~Ronca, E.~F. Kj{\o{o}}nstad, A.~Rubio, and H.~Koch,
  ``Coupled cluster theory for molecular polaritons: Changing ground and
  excited states,'' {\em Phys. Rev. X}, vol.~10, 2020.

\bibitem{Schfer2022}
C.~Sch\"{a}fer, J.~Flick, E.~Ronca, P.~Narang, and A.~Rubio, ``Shining light on
  the microscopic resonant mechanism responsible for cavity-mediated chemical
  reactivity,'' {\em Nat. Commun.}, vol.~13, Dec. 2022.

\bibitem{Wang2019}
X.~Wang, E.~Ronca, and M.~A. Sentef, ``Cavity quantum electrodynamical chern
  insulator: Towards light-induced quantized anomalous hall effect in
  graphene,'' {\em Phys. Rev. B}, vol.~99, p.~235156, June 2019.

\bibitem{Ruggenthaler2014}
M.~Ruggenthaler, J.~Flick, C.~Pellegrini, H.~Appel, I.~V. Tokatly, and
  A.~Rubio, ``Quantum-electrodynamical density-functional theory: Bridging
  quantum optics and electronic-structure theory,'' {\em Phys. Rev. A},
  vol.~90, p.~012508, July 2014.

\bibitem{Flick2017_2}
J.~Flick, H.~Appel, M.~Ruggenthaler, and A.~Rubio, ``Cavity born?oppenheimer
  approximation for correlated electron?nuclear-photon systems,'' {\em J. Chem.
  Theory Comput.}, vol.~13, pp.~1616--1625, Mar. 2017.

\bibitem{Wickramasinghe2025}
S.~Wickramasinghe, A.~Amini, and A.~Mandal, ``On-the-fly cavity-molecular
  dynamics of vibrational polaritons,'' 2025.

\bibitem{Bonini2022}
J.~Bonini and J.~Flick, ``Ab initio linear-response approach to
  vibro-polaritons in the cavity born?oppenheimer approximation,'' {\em J.
  Chem. Theory Comput.}, vol.~18, p.~2764?2773, Apr. 2022.

\bibitem{Bonini2024}
J.~Bonini, I.~Ahmadabadi, and J.~Flick, ``Cavity born?oppenheimer approximation
  for molecules and materials via electric field response,'' {\em J. Chem.
  Phys.}, vol.~161, p.~154104, Oct. 2024.

\bibitem{De2024}
P.~K. De and A.~Jain, ``Exciton energy transfer inside cavity?a benchmark study
  of polaritonic dynamics using the surface hopping method,'' {\em J. Chem.
  Phys.}, vol.~161, p.~054117, Aug. 2024.

\bibitem{Krupp2025}
N.~Krupp, G.~Groenhof, and O.~Vendrell, ``Quantum dynamics simulation of
  exciton-polariton transport,'' {\em Nat. Commun.}, vol.~16, p.~5431, July
  2025.

\bibitem{Hoffmann2019}
N.~M. Hoffmann, C.~Sch\"{a}fer, N.~S\"{a}kkinen, A.~Rubio, H.~Appel, and
  A.~Kelly, ``Benchmarking semiclassical and perturbative methods for real-time
  simulations of cavity-bound emission and interference,'' {\em J. Chem.
  Phys.}, vol.~151, p.~244113, Dec. 2019.

\bibitem{Hoffmann2019_jun}
N.~M. Hoffmann, C.~Sch\"{a}fer, A.~Rubio, A.~Kelly, and H.~Appel, ``Capturing
  vacuum fluctuations and photon correlations in cavity quantum electrodynamics
  with multitrajectory ehrenfest dynamics,'' {\em Phys. Rev. A}, vol.~99,
  p.~063819, June 2019.

\bibitem{Maitra_JCP2020}
N.~M. Hoffmann, L.~Lacombe, A.~Rubio, and N.~T. Maitra, ``Effect of many modes
  on self-polarization and photochemical suppression in cavities,'' {\em J.
  Chem. Phys.}, vol.~153, p.~104103, 2020.

\bibitem{Hu2023}
D.~Hu and P.~Huo, ``Ab initio molecular cavity quantum electrodynamics
  simulations using machine learning models,'' {\em J. Chem. Theory Comput.},
  vol.~19, p.~2353?2368, Mar. 2023.

\bibitem{Sokolovskii2024}
I.~Sokolovskii and G.~Groenhof, ``Non-hermitian molecular dynamics simulations
  of exciton?polaritons in lossy cavities,'' {\em J. Chem. Phys.}, vol.~160,
  p.~092501, Mar. 2024.

\bibitem{Hu2025}
D.~Hu, B.~X.~K. Chng, W.~Ying, and P.~Huo, ``Trajectory-based non-adiabatic
  simulations of the polariton relaxation dynamics,'' {\em J. Chem. Phys.},
  vol.~162, p.~124113, Mar. 2025.

\bibitem{Tichauer2021}
R.~H. Tichauer, J.~Feist, and G.~Groenhof, ``Multi-scale dynamics simulations
  of molecular polaritons: The effect of multiple cavity modes on polariton
  relaxation,'' {\em J. Chem. Phys.}, vol.~154, p.~104112, Mar. 2021.

\bibitem{Vendrell2018}
O.~Vendrell, ``Coherent dynamics in cavity femtochemistry: Application of the
  multi-configuration time-dependent hartree method,'' {\em Chem. Phys.},
  vol.~509, p.~55?65, June 2018.

\bibitem{Li2021}
X.~Li, A.~Mandal, and P.~Huo, ``Cavity frequency-dependent theory for
  vibrational polariton chemistry,'' {\em Nat. Commun.}, vol.~12, p.~1315, Feb.
  2021.

\bibitem{Li2021_2}
T.~E. Li, A.~Nitzan, and J.~E. Subotnik, ``Collective vibrational strong
  coupling effects on molecular vibrational relaxation and energy transfer:
  Numerical insights via cavity molecular dynamics simulations,'' {\em Angew.
  Chem. Int. Ed.}, vol.~60, p.~15533?15540, June 2021.

\bibitem{Rana2023}
B.~Rana, E.~G. Hohenstein, and T.~J. Mart{\'i}nez, ``Simulating the
  excited-state dynamics of polaritons with ab initio multiple spawning,'' {\em
  J. Phys. Chem. A}, vol.~128, pp.~139--151, Dec. 2023.

\bibitem{Yu2022}
Q.~Yu and S.~Hammes-Schiffer, ``Multidimensional quantum dynamical simulation
  of infrared spectra under polaritonic vibrational strong coupling,'' {\em J.
  Phys. Chem. Lett}, vol.~13, p.~11253?11261, Nov. 2022.

\bibitem{Li2022}
T.~E. Li and S.~Hammes-Schiffer, ``Qm/mm modeling of vibrational polariton
  induced energy transfer and chemical dynamics,'' {\em J. Am. Chem. Soc.},
  vol.~145, p.~377?384, Dec. 2022.

\bibitem{Galego2015}
J.~Galego, F.~J. Garcia-Vidal, and J.~Feist, ``Cavity-induced modifications of
  molecular structure in the strong-coupling regime,'' {\em Phys. Rev. X},
  vol.~5, p.~041022, Nov 2015.

\bibitem{Galego2016}
J.~Galego, F.~J. Garcia-Vidal, and J.~Feist, ``Suppressing photochemical
  reactions with quantized light fields,'' {\em Nat. Commun.}, vol.~7,
  p.~13841, Dec. 2016.

\bibitem{Galego2017}
J.~Galego, F.~J. Garcia-Vidal, and J.~Feist, ``Many-molecule reaction triggered
  by a single photon in polaritonic chemistry,'' {\em Phys. Rev. Lett.},
  vol.~119, p.~136001, Sep 2017.

\bibitem{Ghosh2025}
P.~Ghosh, A.~Manjalingal, S.~Wickramasinghe, S.~R. Koshkaki, and A.~Mandal,
  ``Mean-field mixed quantum-classical approach for many-body quantum dynamics
  of exciton polaritons,'' {\em Phys. Rev. B}, vol.~112, Sept. 2025.

\bibitem{Curchod_WIRES2019}
F.~Agostini and B.~F.~E. Curchod, ``Different flavors of nonadiabatic molecular
  dynamics,'' {\em WIREs Comput. Mol. Sci.}, vol.~9, p.~e1417, 2019.

\bibitem{Gross_PRL2010}
A.~Abedi, N.~T. Maitra, and E.~K.~U. Gross, ``Exact factorization of the
  time-dependent electron-nuclear wave function,'' {\em Phys. Rev. Lett.},
  vol.~105, no.~12, p.~123002, 2010.

\bibitem{Agostini_EPJB2021}
F.~Agostini and E.~K.~U. Gross, ``Ultrafast dynamics with the exact
  factorization,'' {\em Eur. Phys. J. B}, vol.~94, p.~179, 2021.

\bibitem{Agostini_PCCP2024}
L.~M. Ibele, E.~S. Gil, E.~V. Arribas, and F.~Agostini, ``Simulations of
  photoinduced processes with the exact factorization: state of the art and
  perspectives,'' {\em Phys. Chem. Chem. Phys.}, vol.~26, pp.~26693--26718,
  2024.

\bibitem{Agostini_CPR2026}
P.~Sch{\"u}rger, S.~Giarrusso, and F.~Agostini, ``Exact factorization of a
  many-body wavefunction beyond the electron-nuclear problem,'' {\em
  arXiv:2602.23914 [physics.chem-ph]}, 2026.

\bibitem{Maitra_EPJB2018}
N.~M. Hoffmann, H.~Appel, A.~Rubio, and N.~Maitra, ``Light-matter interactions
  via the exact factorization approach,'' {\em Euro. Phys. J. B}, vol.~91,
  p.~180, 2018.

\bibitem{Tokatly_EPJB2018}
A.~Abedi, E.~Khosravi, and I.~Tokatly, ``Shedding light on correlated
  electron-photon states using the exact factorization,'' {\em Euro. Phys. J.
  B}, vol.~91, p.~194, 2018.

\bibitem{Maitra_JCP2021}
P.~Martinez, B.~Rosenzweig, N.~M. Hoffmann, L.~Lacombe, and N.~T. Maitra,
  ``Case studies of the time-dependent potential energy surface for dynamics in
  cavities,'' {\em J. Chem. Phys.}, vol.~154, p.~014102, 2021.

\bibitem{Maitra_JCP2022}
B.~Rosenzweig, N.~M. Hoffmann, L.~Lacombe, and N.~T. Maitra, ``Analysis of the
  classical trajectory treatment of photon dynamics for polaritonic
  phenomena,'' {\em J. Chem. Phys.}, vol.~156, p.~054101, 2022.

\bibitem{Maitra_PRL2019}
L.~Lacombe, N.~M. Hoffmann, and N.~T. Maitra, ``Exact potential energy surface
  for molecules in cavities,'' {\em Phys. Rev. Lett.}, vol.~123, p.~083201,
  2019.

\bibitem{Agostini_JCP2024_2}
E.~Sangiogo~Gil, D.~Lauvergnat, and F.~Agostini, ``Exact factorization of the
  photon-electron-nuclear wavefunction: {F}ormulation and coupled-trajectory
  dynamics,'' {\em J. Chem. Phys.}, vol.~161, p.~084112, 2024.

\bibitem{GonZhoRei-EPJB-18}
X.~Gonze, J.~S. Zhou, and L.~Reining, ``Variations on the ``exact
  factorization'' theme,'' {\em Eur. Phys. J. B}, vol.~91, no.~10, p.~224,
  2018.

\bibitem{Maitra_PRL2020}
L.~Lacombe and N.~T. Maitra, ``Embedding via the exact factorization
  approach,'' {\em Phys. Rev. Lett.}, vol.~124, p.~206401, 2020.

\bibitem{Gross_PRL2021}
E.~K. U.~G. Ryan~Requist, ``Fock space embedding theory for strongly correlated
  topological phases,'' {\em Phys. Rev. Lett.}, vol.~127, p.~116401, 2019.

\bibitem{Agostini_CPC2024}
S.~Giarrusso, P.~Gori-Giorgi, and F.~Agostini, ``Electronic vector potential
  from the exact factorization of a complex wavefunction,'' {\em Chem. Phys.
  Chem.}, vol.~25, p.~10.1002/cphc.202400127, 2024.

\bibitem{Agostini_JCP2025_DFT}
S.~Giarrusso and F.~Agostini, ``Modeling the kohn--sham potential for molecular
  dissociation with orbital-independent functionals: A proof of principle,''
  {\em J. Chem. Phys.}, vol.~162, p.~091103, 2025.

\bibitem{CohSteGro-PRB-25}
G.~Cohen, R.~Steinitz-Eliyahu, E.~Gross, S.~Refaely-Abramson, and R.~Requist,
  ``Nonadiabaticity from first principles: {E}xact-factorization approach for
  solids,'' {\em Phys. Rev. B}, vol.~112, no.~7, p.~075102, 2025.

\bibitem{Hunter_IJQC1986}
G.~Hunter, ``The exact one-electron model of molecular structure,'' {\em Int.
  J. Quantum Chem.}, vol.~29, p.~197, 1986.

\bibitem{Schild_JPCL2021}
J.~Koc{\'`}k, E.~Kraisler, and A.~Schild, ``Charge-transfer steps in density
  functional theory from the perspective of the exact electron factorization,''
  {\em J. Phys. Chem. Lett.}, vol.~12, pp.~3204--3209, 2021.

\bibitem{Burghardt_PRL2024_GP}
R.~Martinazzo and I.~Burghardt, ``Dynamics of the molecular geometric phase,''
  {\em Phys. Rev. Lett.}, vol.~132, p.~243002, 2024.

\bibitem{VM23}
E.~Villaseco~Arribas and N.~T. Maitra, ``{Energy-conserving coupled trajectory
  mixed quantum-classical dynamics},'' {\em J. Chem. Phys.}, vol.~158, no.~16,
  p.~161105, 2023.

\bibitem{Franco_JCP2017}
B.~Gu and I.~Franco, ``Partial hydrodynamic representation of quantum molecular
  dynamics,'' {\em J. Chem. Phys.}, vol.~146, p.~194104, 2017.

\bibitem{Suzuki_PRA2016}
Y.~Suzuki and K.~Watanabe, ``Bohmian mechanics in the exact factorization of
  electron-nuclear wave functions,'' {\em Phys. Rev. A}, vol.~94, p.~032517,
  2016.

\bibitem{Scherrer_JCP2015}
A.~Scherrer, F.~Agostini, D.~Sebastiani, E.~K.~U. Gross, and R.~Vuilleumier,
  ``Nuclear velocity perturbation theory for vibrational circular dichroism:
  {A}n approach based on the exact factorization of the electron-nuclear wave
  function,'' {\em J. Chem. Phys.}, vol.~143, no.~7, p.~074106, 2015.

\bibitem{Scherrer_PRX2017}
A.~Scherrer, F.~Agostini, D.~Sebastiani, E.~K.~U. Gross, and R.~Vuilleumier,
  ``On the mass of atoms in molecules: Beyond the born-oppenheimer
  approximation,'' {\em Phys. Rev. X}, vol.~7, p.~031035, 2017.

\bibitem{Blumberger_JCP2025}
A.~Dines and J.~Blumberger, ``Thermal equilibrium in coupled trajectory mixed
  quantum?classical dynamics,'' {\em J. Chem. Phys.}, vol.~163, p.~044116,
  2025.

\bibitem{Nitzan_PNAS2020}
T.~E. Li, J.~E. Subotnik, and A.~Nitzan, ``Cavity molecular dynamics
  simulations of liquid water under vibrational ultrastrong coupling,'' {\em
  Proc. Natl. Acad. Sci. U. S. A.}, vol.~117, pp.~18324--18331, 2020.

\bibitem{Woolley2020}
R.~G. Woolley, ``Power-zienau-woolley representations of nonrelativistic qed
  for atoms and molecules,'' {\em Phys. Rev. Res.}, vol.~2, p.~013206, Feb.
  2020.

\bibitem{Power1959}
A.~Power, S.~Zienau, and H.~S.~W. Massey {\em Philos. Trans. R. Soc. Lond.,
  Ser. A, Math. Phys. Sci.}, vol.~251, p.~427?454, Sept. 1959.

\bibitem{Mandal2020}
A.~Mandal, S.~Montillo~Vega, and P.~Huo, ``Polarized fock states and the
  dynamical casimir effect in molecular cavity quantum electrodynamics,'' {\em
  J. Phys. Chem. Lett.}, vol.~11, p.~9215?9223, Sept. 2020.

\bibitem{Fischer2023}
E.~W. Fischer and P.~Saalfrank, ``Beyond cavity born?oppenheimer: On
  nonadiabatic coupling and effective ground state hamiltonians in
  vibro-polaritonic chemistry,'' {\em J. Chem. Theory Comput.}, vol.~19,
  p.~7215?7229, Oct. 2023.

\bibitem{Agostini_JPCA2022}
L.~M. Ibele, B.~F.~E. Curchod, and F.~Agostini, ``A photochemical reaction in
  different theoretical representations,'' {\em J. Phys. Chem. A}, vol.~126,
  pp.~1263--1281, 2022.

\bibitem{Gross_JCP2015}
F.~Agostini, A.~Abedi, Y.~Suzuki, S.~K. Min, N.~T. Maitra, and E.~K.~U. Gross,
  ``The exact forces on classical nuclei in non-adiabatic charge transfer,''
  {\em J. Chem. Phys.}, vol.~142, no.~8, p.~084303, 2015.

\bibitem{Agostini_JPCL2017}
B.~F.~E. Curchod and F.~Agostini, ``On the dynamics through a conical
  intersection,'' {\em J. Phys. Chem. Lett.}, vol.~8, pp.~831--837, 2017.

\bibitem{Gross_JCTC2016}
F.~Agostini, S.~K. Min, A.~Abedi, and E.~K.~U. Gross, ``Quantum-classical
  non-adiabatic dynamics: {C}oupled- vs. independent-trajectory methods,'' {\em
  J. Chem. Theory Comput.}, vol.~12, no.~5, pp.~2127--2143, 2016.

\bibitem{Ciccotti_EPJB2018}
F.~Agostini, I.~Tavernelli, and G.~Ciccotti, ``Nuclear quantum effects in
  electronic (non)adiabatic dynamics,'' {\em Euro. Phys. J. B}, vol.~91,
  p.~139, 2018.

\bibitem{Ciccotti_JPCA2020}
F.~Talotta, F.~Agostini, and G.~Ciccotti, ``Quantum trajectories for the
  dynamics in the exact factorization framework: a proof-of-principle test,''
  {\em J. Phys. Chem. A}, vol.~124, pp.~6764--6777, 2020.

\bibitem{Gross_JPCL2017}
S.~K. Min, F.~Agostini, I.~Tavernelli, and E.~K.~U. Gross, ``Ab initio
  nonadiabatic dynamics with coupled trajectories: A rigorous approach to
  quantum (de)coherence,'' {\em J. Phys. Chem. Lett.}, vol.~8, pp.~3048--3055,
  2017.

\bibitem{Agostini_JCTC2024}
C.~Pieroni, E.~Sangiogo~Gil, L.~M. Ibele, M.~Persico, G.~Granucci, and
  F.~Agostini, ``Investigating the photodynamics of trans-azobenzene with
  coupled trajectories,'' {\em J. Chem. Theory Comput.}, vol.~20, pp.~580--596,
  2024.

\bibitem{Scribano_JCTC2022}
L.~Dupuy, F.~Talotta, F.~Agostini, D.~Lauvergnat, B.~Poirier, and Y.~Scribano,
  ``Adiabatic and nonadiabatic dynamics with interacting quantum
  trajectories,'' {\em J. Chem. Theory Comput.}, vol.~18, pp.~6447--6462, 2022.

\bibitem{Rassolov_JCTC2023}
S.~Garashchuk, J.~Stetzler, and V.~Rassolov, ``Factorized electron?nuclear
  dynamics with an effective complex potential,'' {\em J. Chem. Theory
  Comput.}, vol.~19, pp.~1393--1408, 2023.

\bibitem{Ananth_JCP2013}
N.~Ananth, ``Mapping variable ring polymer molecular dynamics: {A}
  path-integral based method for nonadiabatic processes,'' {\em J. Chem.
  Phys.}, vol.~139, p.~124102, 2013.

\bibitem{Huo_JCP2017}
S.~N. Chowdhury and P.~Huo, ``Coherent state mapping ring polymer molecular
  dynamics for non-adiabatic quantum propagations,'' {\em J. Chem. Phys.},
  vol.~147, p.~214109, 2017.

\bibitem{Schfer2018}
C.~Sch\"{a}fer, M.~Ruggenthaler, and A.~Rubio, ``Ab initiononrelativistic
  quantum electrodynamics: Bridging quantum chemistry and quantum optics from
  weak to strong coupling,'' {\em Phys. Rev. A}, vol.~98, p.~043801, Oct. 2018.

\bibitem{Fregoni2020}
J.~Fregoni, S.~Corni, M.~Persico, and G.~Granucci, ``Photochemistry in the
  strong coupling regime: A trajectory surface hopping scheme,'' {\em J.
  Comput. Chem.}, vol.~41, p.~2033?2044, July 2020.

\bibitem{Gross_PRL2015}
S.~K. Min, F.~Agostini, and E.~K.~U. Gross, ``Coupled-trajectory
  quantum-classical approach to electronic decoherence in nonadiabatic
  processes,'' {\em Phys. Rev. Lett.}, vol.~115, no.~7, p.~073001, 2015.

\bibitem{TULLY1998}
J.~C. TULLY, ``Mixed quantum-classical dynamics: mean-field and
  surface-hopping,'' in {\em Classical and Quantum Dynamics in Condensed Phase
  Simulations}, p.~489?514, WORLD SCIENTIFIC, June 1998.

\bibitem{Tully_JCP1990}
J.~C. Tully, ``Molecular dynamics with electronic transitions,'' {\em J. Chem.
  Phys.}, vol.~93, p.~1061, 1990.

\bibitem{Agostini_JCTC2020_1}
F.~Talotta, S.~Morisset, N.~Rougeau, D.~Lauvergnat, and F.~Agostini, ``Internal
  conversion and intersystem crossing with the exact factorization,'' {\em J.
  Chem Theory Comput.}, vol.~16, pp.~4833--4848, 2020.

\bibitem{IBELE2024188}
L.~M. Ibele, C.~Pieroni, F.~Talotta, B.~F. Curchod, D.~Lauvergnat, and
  F.~Agostini, ``Exact factorization of the electron-nuclear wavefunction:
  Fundamentals and algorithms,'' in {\em Comprehensive Computational Chemistry
  (First Edition)} (M.~Y{\'a}{\~n}ez and R.~J. Boyd, eds.), pp.~188--211,
  Oxford: Elsevier, first edition~ed., 2024.

\bibitem{QML}
D.~Lauvergnat, ``{QuantumModelLib},'' accessed in June 2024.
\newblock \url{https://github.com/lauvergn/QuantumModelLib}.

\bibitem{Steiger_JCP1982}
M.~D. Feit, F.~A. {Fleck~Jr.}, , and A.~Steiger, ``Solution of the
  {S}chr{\"o}dinger equation by a spectral method,'' {\em J. Comput. Phys.},
  vol.~47, p.~412, 1982.

\bibitem{GCTMQC}
F.~Agostini, E.~Marsili, F.~Talotta, C.~Pieroni, E.~Villaseco~Arribas, L.~M.
  Ibele, and E.~Sangiogo~Gil, ``{G-CTMQC},'' accessed in June 2024.
\newblock \url{https://gitlab.com/agostini.work/g-ctmqc}.

\bibitem{Hoffmann2020}
N.~M. Hoffmann, L.~Lacombe, A.~Rubio, and N.~T. Maitra, ``Effect of many modes
  on self-polarization and photochemical suppression in cavities,'' {\em J.
  Chem. Phys.}, vol.~153, p.~104103, Sept. 2020.

\bibitem{AgostiniEich_JCP2016}
F.~G. Eich and F.~Agostini, ``The adiabatic limit of the exact factorization of
  the electron-nuclear wave function,'' {\em J. Chem. Phys.}, vol.~145,
  p.~054110, 2016.

\bibitem{Agostini_JCTC2021}
C.~Pieroni and F.~Agostini, ``Nonadiabatic dynamics with coupled
  trajectories,'' {\em J. Chem. Theory Comput.}, vol.~17, p.~5969, 2021.

\end{thebibliography}

\end{document}